\documentclass[twocolumn,prd,a4paper,final,superscriptaddress,longbibliography]{revtex4-1}

\usepackage{amsmath,amssymb,amsfonts}
\usepackage[dvipdfmx]{graphicx}
\usepackage[dvipdfmx]{hyperref}
\usepackage{bbold}
\usepackage[T1]{fontenc}
\usepackage{epsfig}
\usepackage{epstopdf}
\usepackage{bbold}
\usepackage{comment}
\usepackage{bm}
\usepackage{here}

\usepackage{natbib}

\graphicspath{{./fig/}}

\begin{document}
	
	\title{Phase analyses for compact, charged boson stars and shells harboring black holes 
	in the $\mathbb{C}P^N$ nonlinear sigma model}
	
	\author{Nobuyuki Sawado}
	\email{sawadoph@rs.tus.ac.jp}
	\affiliation{Department of Physics, Tokyo University of Science, Noda, Chiba 278-8510, Japan}

	\author{Shota Yanai}
	\email{phyana0513@gmail.com}
	\affiliation{Department of Physics, Tokyo University of Science, Noda, Chiba 278-8510, Japan}

	\vspace{.5 in}
	\small

	\date{\today}
	
	\begin{abstract}
	Phase diagrams of the boson stars and shells of the $U(1)$ gauged $\mathbb{C}P^N$ nonlinear sigma model are 
	studied. The solutions of the model exhibit both the ball- and the shell-shaped charge density depending on $N$. 
	There appear four independent regions of the solutions which are essentially caused from the coexistence of 
	electromagnetism and gravity. We examine several phase diagrams of the boson stars and the shells 
	and discuss what and how the regions are emerged. 
	A coupling with gravity allows for harboring of the charged black holes for the $Q$-shell 
	solutions. Some solutions are strongly affected by the presence of the black holes and
	they allow to be smoothly connected. 
	As a result, the regions are integrated by the harboring black holes. 	
	\end{abstract}
	
	\pacs{}

	\maketitle 
	\section{Introduction}
	\label{Intro}
	
	A complex scalar field theory with some self-interactions has stationary
	soliton solutions called $Q$-balls~\cite{Friedberg:1976me,Coleman:1985ki,Leese:1991hr,Loiko:2018mhb}. 
	$Q$-balls have attracted much attention in the studies of 
	evolution of the early Universe~\cite{Friedberg:1986tq,Lee:1986ts}.
	In supersymmetric extensions of the standard model, $Q$-balls appear as
	the scalar superpartners of baryons or leptons forming coherent states with 
	baryon or lepton number. They may survive as a major ingredient of dark matter
	~\cite{Kusenko:1997zq,Kusenko:1997si,Kusenko:1997vp}.
	
	Analysis in this paper is based on the $Q$-ball solutions of the $\mathbb{C}P^N$ nonlinear sigma model 
	which is defined by the Lagrangian density~\cite{Klimas:2017eft}
	\begin{align}
	{\cal L}=-\frac{M^2}{2}\mathrm{Tr}(X^{-1}\partial_\mu X)^2-\mu^2V(X),
	\label{lag0}
	\end{align}
	where the `V-shaped' potential
	\begin{align}
	V(X)=\frac{1}{2}[\mathrm{Tr}(I-X)]^{1/2}
	\end{align} 
	is employed in order to obtain the compact solutions. The behavior of fields at the outer border of compacton implies $X\to I$. 
	The coupling constants $M$ and $\mu$ have dimensions of $(\mathrm{length})^{-1}$ and $(\mathrm{length})^{-2}$, respectively. 
	The principal variable $X$ successfully parametrizes the coset space
	 $SU(N+1)/U(N)\sim \mathbb{C}P^N$. It is parametrized by complex fields $u_i$ in the following way
	\begin{align}
	X(g)=
	\left(\begin{array}{cc}
	I_{N\times N} & 0 \\
	0 & -1 
	\end{array}\right)+
	\frac{2}{\vartheta^2}
	\left(\begin{array}{cc}
	-u\otimes u^\dagger & iu \\
	iu^\dagger & 1  
	\end{array}\right)
	\label{principalu}
	\end{align}
	where $\vartheta:=\sqrt{1+u^\dagger\cdot u}$. 		
	Thus the $\mathbb{C}P^N$ Lagrangian of the model (\ref{action}) 
	takes the form
	\begin{align}
	{\cal L}_{\mathbb{C}P^N}=-M^2g^{\mu\nu}\tilde{\tau}_{\nu\mu}-\mu^2 V
	\label{lagrangianu}
	\end{align}
	where
	\begin{align}
	\tilde{\tau}_{\nu\mu}=-\frac{4}{\vartheta^4}\partial_\mu u^\dagger\cdot \Delta^2\cdot \partial_\nu u,~~
	\Delta_{ij}^2:=\vartheta^2\delta_{ij}-u_iu_j^*\,.
	\end{align}
	The model possesses the compactons~\cite{Klimas:2017eft}.  
	Compactons are field configurations that exist on finite size supports and outside this support, the field is 
	identically zero. For example, the signum-Gordon model, i.e., the scalar field model with standard
	kinetic terms and V-shaped potential gives rise to such solutions~\cite{Arodz:2008jk,Arodz:2008nm}. 
	
	In last few years, we made some efforts in the study of compact boson stars corresponding to the model~(\ref{lag0})
	~\cite{Klimas:2018ywv, Yanai:2019wpv, Sawado:2020ncc}.			
	The boson stars are the gravitating objects of such $Q$-balls. There are a large number of papers concerning the 
	boson stars
	~\cite{Lee:1991ax, Friedberg:1986tq, Jetzer:1991jr, Liebling:2012fv, Kleihaus:2009kr,
	Kleihaus:2010ep,Kumar:2014kna,Kumar:2015sia,Kumar:2016sxx}.
	The gravitating boson shells can harbor a Schwarzschild and a Reissner-Nordstr\"om type black hole. 
	The harbor is a solution that is as follows. 
	When in the center of shell is a localized massive body, such as the Schwarzschild-like black hole,  
	we set the event horizon in the interior part of the shell and solve the equations from the event horizon 
	to the outer region. Such solutions are called harbor~\cite{Kleihaus:2010ep}. 
	Since the black hole is surrounded by a shell of scalar fields, such
	fields outside of the event horizon may be interpreted as a scalar hair. 	
	The excited boson stars are very important not only from theoretical interest but also 
	for astrophysical observations~\cite{Brihaye:2008cg,Bernal:2009zy,Collodel:2017biu,Alcubierre:2018ahf,Alcubierre:2019qnh}. 
	The multistate boson stars, which are superposed ground and excited state boson star solutions, are considered 
	for obtaining realistic rotation curves of spiral galaxies~\cite{Bernal:2009zy}. 
	In~\cite{Alcubierre:2019qnh, Jaramillo:2020rsv}, the stability analysis for their multistate solutions is extensively studied. 
	Our model is a different type of the multistate boson stars on target space $\mathbb{C}P^N$. 
	For the $U(1)$ gauged model~\cite{Sawado:2020ncc}, we observed the signal of the bifurcation 
	and the domain structure in the solutions. These boson shells also harbor a Schwarzschild and a Reissner-Nordstr\"om black hole.
	In this paper, we discuss several novel results for the phase structure of our $U(1)$ gauged 
	gravitating boson stars and shells. The extensive analysis of a single scalar model has already been done in many 
	literatures~\cite{Kleihaus:2009kr,Kleihaus:2010ep,Kumar:2014kna,Kumar:2015sia,Kumar:2016sxx,Kumar:2019dbi}.
	In our model, we observe several bifurcations of the solutions which previously were not known and   
	give us the deeper insights for the interactions forming the boson stars/shells and also for the 
	property of the harbor of the black holes. 
	
	The paper is organized as follows. In Sec.II we shall describe the model,
	coupled to the gravity. Ansatz for the parametrization of the $\mathbb{C}P^N$ field 
	is given in this section. 	
	Section III is the phase diagrams for the $\mathbb{C}P^1$ boson star and shell. 
	We give the boson shell solutions of $N=11$ in Sec.IV. Further discussion 
	for the interpretation of the phase diagram as well as the property of the 
	harbor is discussed in Sec.V. 
	Conclusions and remarks are presented in the last section.

	\section{The model}
	\label{model}
	
	In \cite{Klimas:2018ywv, Yanai:2019wpv, Sawado:2020ncc}, we described formalism of the $\mathbb{C}P^N$ model 
	in flat space-time and also the model of gravitating $Q$-balls and -shells in detail. 
	Here, we briefly review the formalism. 
	The action of self-gravitating complex fields $u_i$ coupled to Einstein gravity has the form
	\begin{align}
	&S=\int d^4x\sqrt{-g}\biggl[
	\frac{R}{16\pi G}-\frac{1}{4}g^{\mu\lambda}g^{\nu\sigma}F_{\mu\nu}F_{\lambda\sigma} \nonumber \\
	&\hspace{4cm}-M^2g^{\mu\nu}\tau_{\nu\mu}-\mu^2 V
	\biggr]\,, 
	\label{action} \\
	&\tau_{\nu\mu}=-\frac{4}{\vartheta^4}D_\mu u^\dagger\cdot \Delta^2\cdot D_\nu u,~~
	\Delta_{ij}^2:=\vartheta^2\delta_{ij}-u_iu_j^*\,.
	\end{align}
	where $G$ is Newton's gravitational constant. $F_{\mu\nu}$ is the standard electromagnetic field tensor 
	and the complex fields $u_i$ also are minimally coupled to the Abelian gauge fields $A_\mu$ through
	$D_\mu=\partial_\mu-ieA_\mu$.
	The variation of the action with respect to the metric leads to Einstein's equations
	\begin{align}
	G_{\mu\nu}=8\pi GT_{\mu\nu},\quad{\rm where}\quad G_{\mu\nu}\equiv R_{\mu\nu}-\frac{1}{2}g_{\mu\nu}R
	\label{einstein_formal}
	\end{align}
	where the stress-energy tensor reads
	\begin{align}
	T_{\mu\nu}=g_{\mu\nu}(M^2g^{\lambda\sigma}\tau_{\sigma\lambda}+\frac{1}{4}g^{\lambda\sigma}g^{\eta\delta}F_{\lambda\eta}F_{\sigma\delta}+\mu^2V)
	\nonumber \\
	-2M^2\tau_{\nu\mu}-g^{\lambda\sigma}F_{\mu\lambda}F_{\nu\sigma}\,.
	\label{stress_formal}
	\end{align}
	The field equations of the complex fields are obtained by variation of the Lagrangian with respect to $u_i^*$
	\begin{align}
	&\frac{1}{\sqrt{-g}}D_\mu (\sqrt{-g} D^\mu u_i)-\frac{2}{\vartheta^2}(u^\dagger\cdot D^\mu u)D_\mu u_i \nonumber \\
	&+\frac{\mu^2}{4M^2}\vartheta^2\sum_{k=1}^{N}\biggl[(\delta_{ik}+u_iu_k^*)\frac{\partial V}{\partial u_k^*}\biggr]=0\,.
	\label{CPNeq}
	\end{align}
	The Maxwell's equations read 
	\begin{align}
	\frac{1}{\sqrt{-g}}\partial_\nu (\sqrt{-g}F^{\nu\mu})=\frac{4ie}{\vartheta^4}M^2(u^\dagger\cdot D^\mu u-D^\mu u^\dagger\cdot u)\,.
	\label{Maxwell}
	\end{align}
	
	It is convenient to introduce the dimensionless coordinates 
	\begin{align}
	x_\mu \to \frac{\mu}{M}x_\mu
	\end{align}
	and also $A_\mu \to \mu/M A_\mu$. 
	We also restrict $N$ to be odd, i.e., $N:=2n+1$.  For solutions with vanishing magnetic field 
	the ansatz has the form
	\begin{align}
	&u_m(t,r,\theta,\varphi)=\sqrt{\frac{4\pi}{2n+1}}f(r)Y_{nm}(\theta,\varphi)e^{i\omega t}\,, 
	\label{ansatzcpn}\\
	&A_\mu(t,r,\theta,\varphi)dx^\mu=A_t(r)dt
	\label{ansatzgauge}
	\end{align}
	and it allows for reduction of the partial differential equations to the system of radial ordinary 
	differential equations.
	$Y_{nm}, -n\leq m \leq n$ are the standard spherical harmonics and $f(r)$ is the matter profile function. 
	Each $2n+1$ field $u=(u_m)=(u_{-n},u_{-n+1},\cdots,u_{n-1},u_n)$ is associated with one of $2n+1$
	spherical harmonics for given $n$. 
	The relation 
	$\sum_{m=-n}^n Y_{nm}^*(\theta,\varphi)Y_{nm}(\theta,\varphi)=\dfrac{2n+1}{4\pi}$
	is very useful for obtaining an explicit form of many inner products. 
	We introduce a new gauge field concerning the gauge field for convenience,
	\begin{align}
	b(r):=\omega-eA_t(r)\,.
	\label{ansatzgaugew}
	\end{align}
	Using the ansatz, we find the dimensionless Lagrangian of the $\mathbb{C}P^N$ model in the form 
	\begin{align}
	&\tilde{\mathcal{L}}_{\mathbb{C}P^N}
	=-\frac{\kappa}{4}g^{\mu\lambda}g^{\nu\sigma}F_{\mu\nu}F_{\lambda\sigma}-g^{\nu\mu}\tau_{\nu\mu}-V
	\nonumber \\
	&=\frac{\kappa b'^2}{2A^2e^2}+\frac{4b^2f^2}{A^2C(1+f^2)^2}-\frac{4Cf'^2}{(1+f^2)^2}-\frac{4n(n+1)f^2}{r^2(1+f^2)}-V
	\label{effectivelag}
	\end{align}  
	where we introduced the dimensionless constant $\kappa:=\mu^2/M^4$ for convenience.
	
	For the ansatz (\ref{ansatzcpn})$-$(\ref{ansatzgaugew}),  
	a suitable form of line element is the standard spherically symmetric Schwarzschild-like coordinates defined by
	\begin{align}
	ds^2 &= g_{\mu \nu}dx^{\mu}dx^{\nu} \nonumber \\
	&=A^2(r)C(r)dt^2 - \frac{1}{C(r)}dr^2 - r^2(d\theta^2 +\sin^2 \theta d\varphi^2)\,.
	\label{metric}
	\end{align}    
	The equations of motion of $A(r),C(r)$ read
	\begin{align}
	&A'=4\alpha r \Bigl[\frac{b^2 f^2}{A^2 C^2(1+f^2)^2}+\frac{f'^2}{(1+f^2)^2} \Bigr]\,, 
	\label{eq:N}\\
	&C'=\frac{1-C}{r}
	\nonumber \\
	&-\alpha r\Bigl[\frac{4b^2 f^2}{A^2 C(1+f^2)^2}+\frac{4Cf'^2}{(1+f^2)^2}+\frac{4n(n+1)f^2}{(1+f^2)r^2}
	\nonumber \\
	&\hspace{1cm}+\frac{\kappa b'^2}{2A^2e^2} +\frac{f}{\sqrt{1+f^2}}\Bigr]\,,
	\label{eq:C}
	\end{align}
	where $\alpha:=8\pi G\mu^2$ is a dimensionless coupling constant concerning to the gravity.
	Plugging the ansatz (\ref{ansatzcpn})-(\ref{ansatzgaugew}) into the matter field equation
	 (\ref{CPNeq}) and the Maxwell's equations (\ref{Maxwell}), we have 
	\begin{align}
	&Cf'' +C'f'+\frac{A'Cf'}{A} +\frac{2C}{r}f' - \frac{n(n+1)f}{r^2}
	\nonumber \\
	&\hspace{1cm}+\frac{(1-f^2)b^2f}{A^2C(1+f^2)}-\frac{2Cff'^2}{(1+f^2)} -\frac{1}{8}\sqrt{1+f^2} =0\,, 
	\label{eq:f}\\
	&\kappa b''+\frac{2r'A-A'r}{Ar}\kappa b' - \frac{8e^2}{C}\frac{bf^2}{(1+f^2)^2} =0\,.
	\label{eq:b}
	\end{align}
	Thus, we solve a system of four coupled equations (\ref{eq:N})$-$(\ref{eq:b}) varying the parameters $\alpha$
	with fixed $\kappa,e$ (in this paper we simply set $\kappa=e=1$).
	
	The dimensionless Hamiltonian of the model is easily obtained,
	\begin{align}
	\mathcal{H}_{\mathbb{C}P^N}
	&=\frac{4 b^2 f^2}{A^2C(1+f^2)^2}+\frac{\kappa b'^2}{2A^2e^2}+\frac{4Cf'^2}{(1+f^2)^2}
	\nonumber \\
	&\hspace{1cm}+\frac{4n(n+1)f^2}{r^2(1+f^2)}+V\,.
	\end{align}
	The total energy is thus given by
	\begin{align}
	&E=4\pi\int r^2dr
	\biggl[
	\frac{\kappa b'^2}{2Ae^2}+\frac{4b^2f^2}{AC(1+f^2)^2}
	\nonumber \\
	&\hspace{1cm}+\frac{4ACf'^2}{(1+f^2)^2}+\frac{4An(n+1)f^2}{r^2(1+f^2)}+AV\biggr]\,.
	\label{energy}
	\end{align}

	The action (\ref{action}) with the covariant derivative is invariant under the
	following local $U(1)^N$ symmetry
	\begin{align}
	&A_\mu(x) \to A_\mu(x)+e^{-1}\partial_\mu\Lambda(x) \nonumber \\
	&u_i\to \exp[iq_i \Lambda(x)]u_i,~~~~i=1,\cdots,N\,,
	\label{gtransformation}
	\end{align}	
	where $q_i$ are some real numbers. 
	The following Noether current is associated with the invariance of the action (\ref{action})  
	under transformations (\ref{gtransformation}):
	\begin{align}
	J^{(i)}_\mu=-\frac{4M^2i}{\vartheta^4}\sum_{j=1}^N[u_i^* \Delta_{ij}^2 D_\mu u_j-D_\mu u_j^* \Delta_{ji}^2u_i]\,.
	\end{align}
	Using the ansatz (\ref{ansatzcpn}) and (\ref{ansatzgauge}) we find the following form of the Noether currents
	\begin{align}
	J_t^{(m)}= \frac{(n-m)!}{(n+m)!}\frac{8bf^2}{(1+f^2)^2}\bigl(P^m_n(\cos\theta)\bigr)^2\,,
	\label{current0} \\
	J_\varphi^{(m)}=\frac{(n-m)!}{(n+m)!}\frac{8mf^2}{(1+f^2)^2}\bigl(P^m_n(\cos\theta)\bigr)^2
	\label{currentp}
	\end{align}
	and $J_r^{(m)}=J_\theta^{(m)}=0$ for $m=-n,-n+1,\cdots,n-1,n$.
	The conservation of currents is explicit after writing the continuity equation in the form
	\begin{align}
	\frac{1}{\sqrt{-g}}\partial_\mu (\sqrt{-g}g^{\mu\nu}J_\nu^{(m)})=\frac{1}{A^2C}\partial_t J_t^{(m)}
	+\frac{1}{r^2\sin^2\theta}\partial_\varphi J_\varphi^{(m)}=0
	\end{align}	
	Therefore, the corresponding Noether charge is
	\begin{align}
	Q^{(m)}&=\frac{1}{2}\int_{\mathbb{R}^3} d^3x \sqrt{-g} \frac{1}{A^2C}J^{(m)}_t(x) \nonumber \\
	&=\frac{16\pi}{2n+1}\int r^2dr\frac{bf^2}{AC(1+f^2)^2}\,.
	\end{align}
	Owing to our ansatz, the charge does not depend on index $m$, which means the symmetry of the solutions 
	is reduced to $U(1)$. However, we shall keep the index for completeness.

	The boundary conditions at the border(s) of the compacton are examined  
 	in terms of expansions.  
	At the origin, the solutions are represented by series
	\begin{align}
	&f(r)=\sum_{k=0}^\infty f_kr^k\,,~~~~
	b(r)=\sum_{k=0}^\infty b_kr^k\,, \nonumber \\
	&A(r)=\sum_{k=0}^\infty A_kr^k\,,~~~~
	C(r)=\sum_{k=-2}^\infty C_kr^k\,.
	\end{align}
	After substituting these expressions into equations (\ref{eq:N})$-$(\ref{eq:b}), 
	one requires vanishing of equations in all orders of expansion. It allows us to determinate the coefficients of expansion. 
	The form is given for each value of parameter $n$.
	For $n=0$, it reads
	\begin{align}
	&f(r)=f_0+\frac{1}{48}\biggl(\sqrt{1+f_0^2}-\frac{8f_0(1-f_0^2)b_0^2}{A_0^2(1+f_0^2)}\biggr)r^2+O(r^4)\,, \nonumber \\
	&b(r)=b_0 + \frac{4 e^2 b_0^2 f_0^2}{3(1+f_0^2)^2} r^2 + O(r^4)\,, \\
	&A(r)=A_0+\frac{2\alpha f_0^2b_0^2}{A_0(1+f_0^2)^2}r^2+O(r^4)\,,\label{expansion0} \\
	&C(r)=1-\frac{\alpha}{3}\biggl(\frac{f_0}{\sqrt{1+f_0}}+\frac{4f_0^2b_0^2}{A_0^2(1+f_0^2)^2}\biggr)r^2+O(r^4) \nonumber
	\end{align}
	where $f_0$, $ b_0$ and $A_0$ are free parameters.
	For $n=1$, we obtain
	\begin{align}
	&f(r)=f_1r+\frac{1}{32}r^2+\frac{1}{10}\biggl(2f_1^3(1+6\alpha)-\frac{f_1b_0^2}{A_0^2}\biggr)r^3+O(r^4)\,, \nonumber \\
	&b(r)=b_0 +\frac{2}{5}e^2 f_1^2 b_0 r^4 + O(r^5)\,, \\
	&A(r)=A_0+\alpha A_0f_1^2r^2+\frac{1}{6}\alpha A_0f_1r^3+O(r^4)\,,\label{expansion1} \\
	&C(r)=1-4\alpha f_1^2r^2 -\frac{\alpha f_1}{2}r^3+ O(r^4) \nonumber
	\end{align}
	with free parameters $f_1$, $b_0$, and $A_0$. 

	For $n\geqq 2$, we have no nontrivial solutions at the vicinity of the origin $r=0$, and
	the solution has to be identically zero. 
	In order to get nontrivial solution, we consider a possibility that the solution does not 
	vanish only inside the shell having radial support $r\in (R_{\rm in}, R_{\rm out})$. Solutions of this
	kind are called $Q$-shells. 
	We study expansion at the sphere with an inner or outer radius. 
	Expansions at both borders of the compacton are very similar. 
	We impose the following boundary conditions at the compacton radius $r=R~(\equiv R_{\rm in},R_{\rm out})$:
	\begin{align}
	f(R)=0,~~f'(R)=0,~~A(R)=1\,.
	\label{compacton}
	\end{align}
	The functions $f(r)$, $b(r)$, $A(r)$ and $C(r)$ are represented by series 
	\begin{align}
	&f(r)=\sum_{k=2}^\infty F_k(R-r)^k,~~b(r)=\sum_{k=0}^\infty B_k(R-r)^k,\nonumber \\
	&A(r)=\sum_{k=0}^\infty A_k(R-r)^k,~~ C(r)=\sum_{k=-2}^\infty C_k(R-r)^k.
	\end{align}
	First few terms have the form 
	\begin{align}
	&f(r)=\frac{R}{16C_0}(R-r)^2+\frac{R}{24C_0^2}(R-r)^3 +O((R-r)^4)\,, 
	\nonumber \\
	&b(r)=B_0 + B_1(R-r)-\frac{B_1}{R}(R-r)^2 + \frac{B_1}{3R^2}(R-r)^3 
	\nonumber \\
	&\hspace{5cm}+O((R-r)^4)\,, \nonumber \\
	&A(r)=A_0-\frac{\alpha R}{48C_0^2}(R-r)^3+O((R-r)^4)\,, \\
	&C(r)=C_0+\frac{1-C_0}{R}(R-r) 
	\nonumber \\
	&\hspace{1cm} +\biggl\{
	\biggl(C_0- 1 \biggr)\frac{1}{R_0^2}-\frac{5\alpha B_1^2}{4A_0^2e^2}
	\biggr\}(R-r)^2  \nonumber \\
	&\hspace{5cm} + O((R-r)^3) \,. \nonumber 
	\end{align}

\begin{figure*}[t]
  \begin{center}
    \includegraphics[width=80mm]{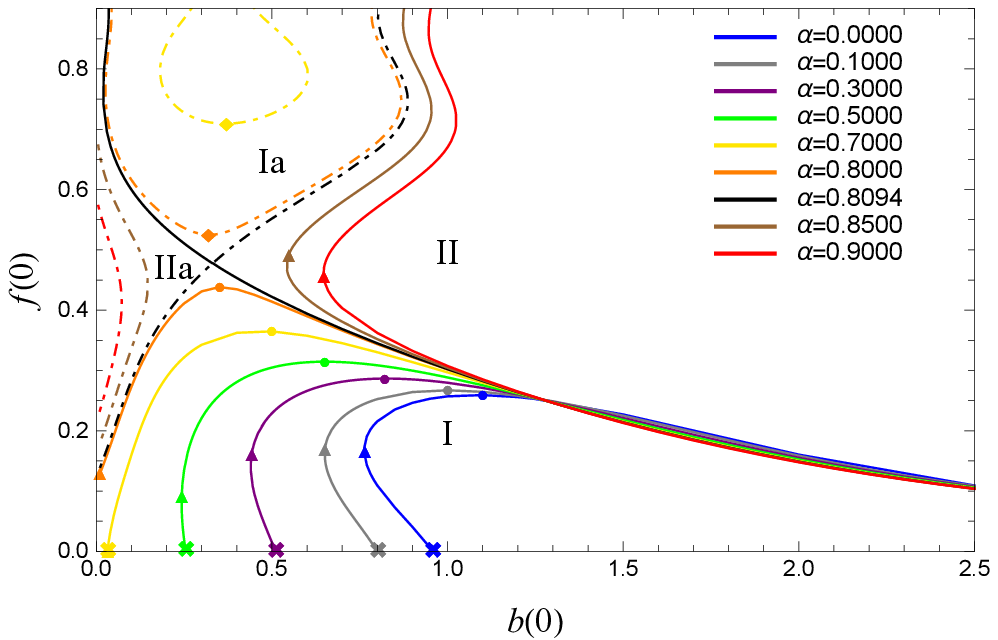}~~
    \includegraphics[width=80mm]{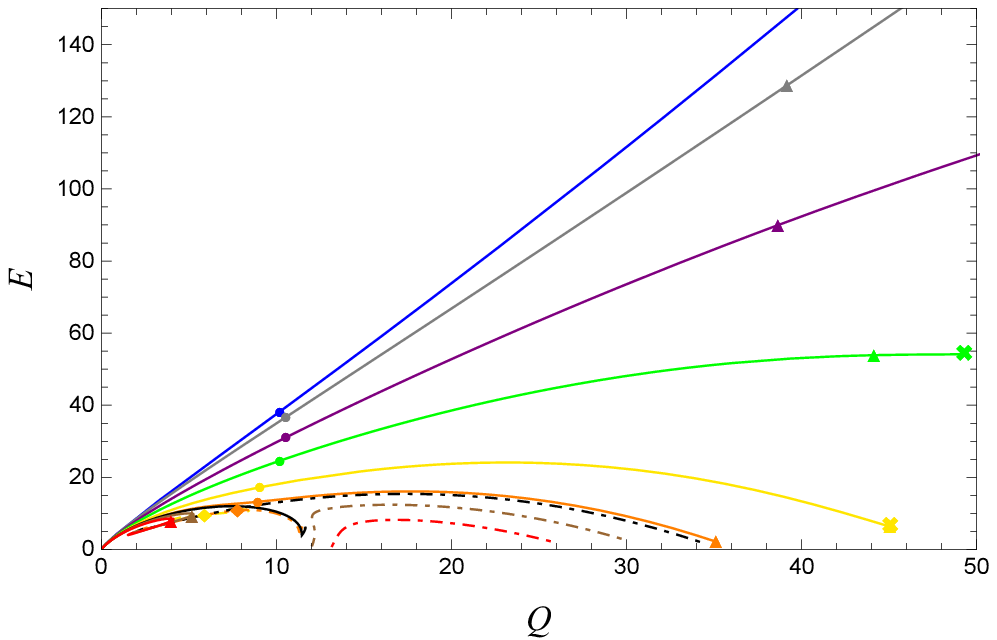}\\
(a)\hspace{8cm} (b)\\
\vspace{5mm}
 
    \includegraphics[width=80mm]{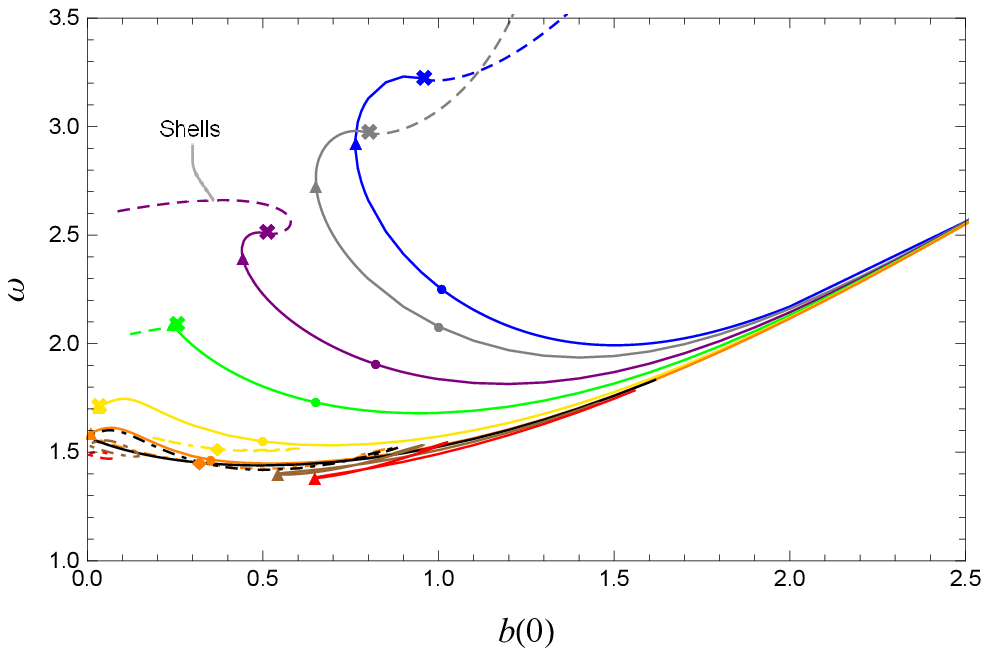}~~
    \includegraphics[width=80mm]{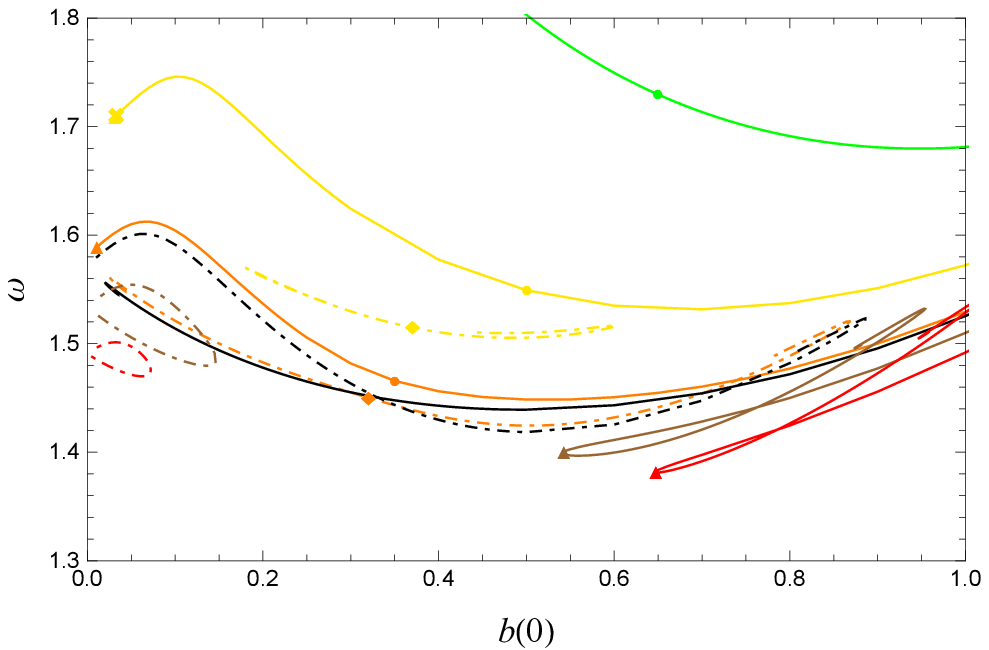} \\
(c)\hspace{8cm} (d)\\ 
\vspace{5mm}

   \includegraphics[width=80mm]{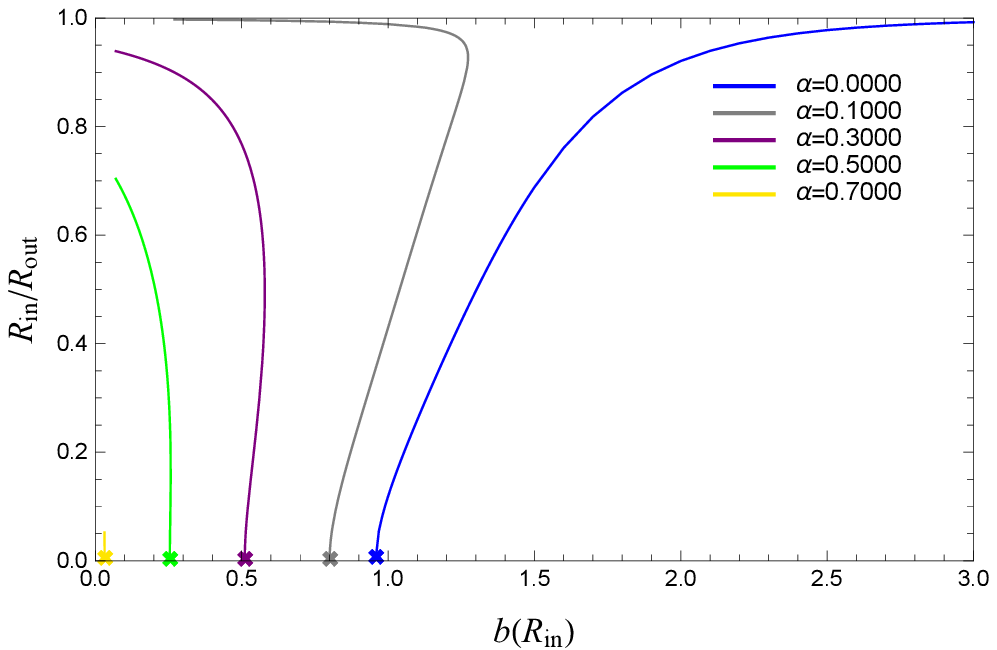}~~
   \includegraphics[width=80mm]{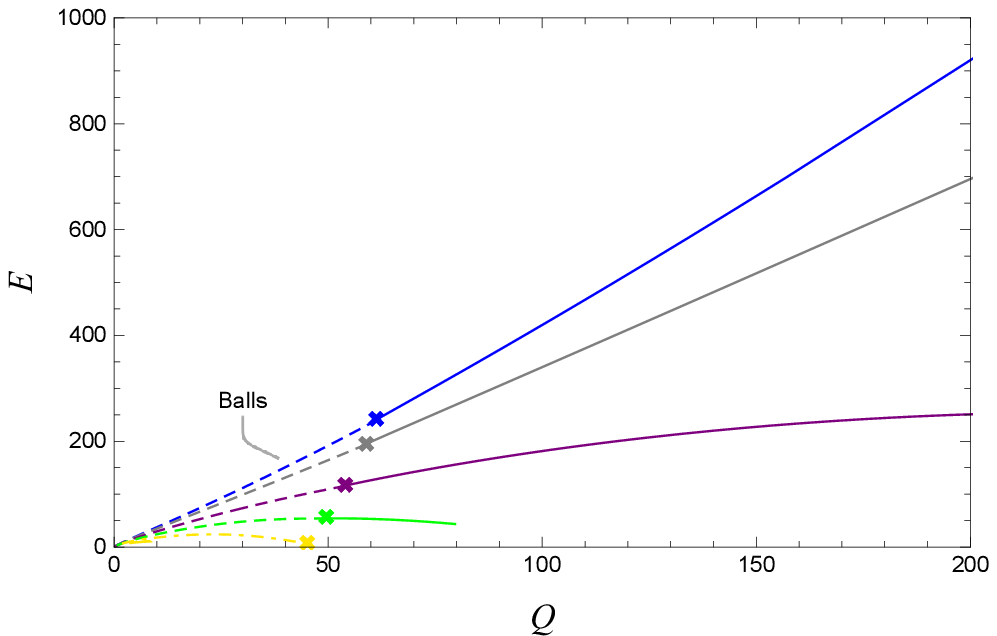}\\
(e)\hspace{8cm} (f)\\

   \caption{\label{CP1phase}  The $\mathbb{C}P^{1}$ boson star and shell.
For the boson stars:  
(a): the phase diagram with the shooting parameters $f(0), b(0)$, the value of the matter profile function $f(r)$ and 
the gauge field function $b(r)$ at the origin, 
(b):~the relation between $E$ and $Q$, 
(c): a relation between the $b(0)$ and frequency $\omega$,  
(d): the same as (c)  but the plots 
$0.0 \le b(0)\le 1.0, 1.3\le \omega \le 1.8$ are enlarged. 
For the boson shells: 
(e): the phase diagram where the ratio of inner and outer shell radii $R_{\rm in}/R_{\rm out}$
and the gauge field at the inner radius $b(R_{\rm in})$,  
(f):~the relation between $E$ and $Q$. The dashed lines are the corresponding ball solutions of the region I. 
}
  \end{center}
	\end{figure*}

\begin{figure*}[htbp]
  \begin{center}
    \includegraphics[width=80mm]{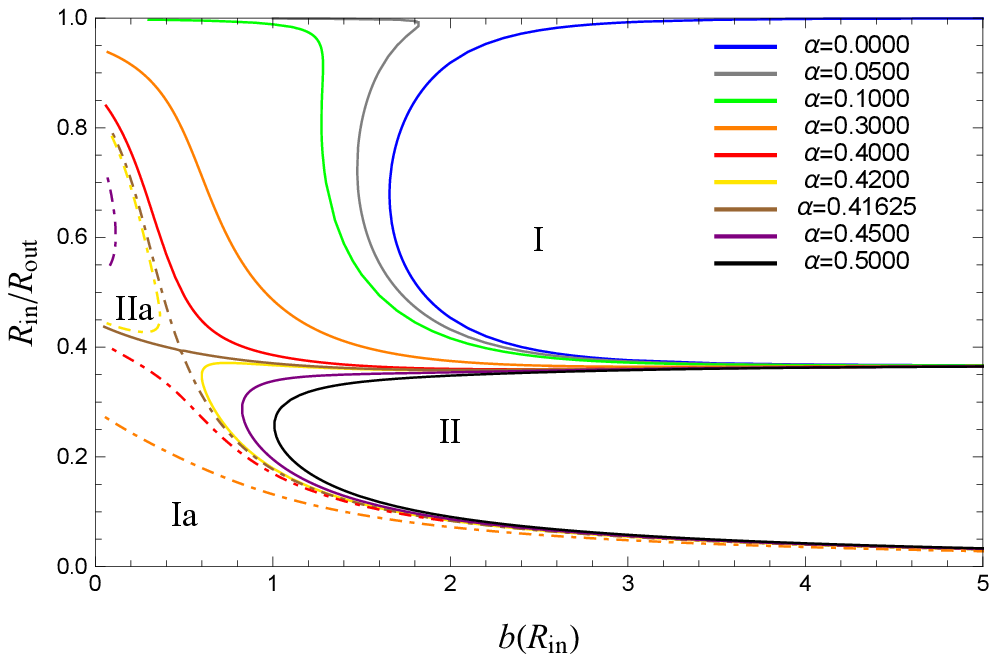}~~
    \includegraphics[width=80mm]{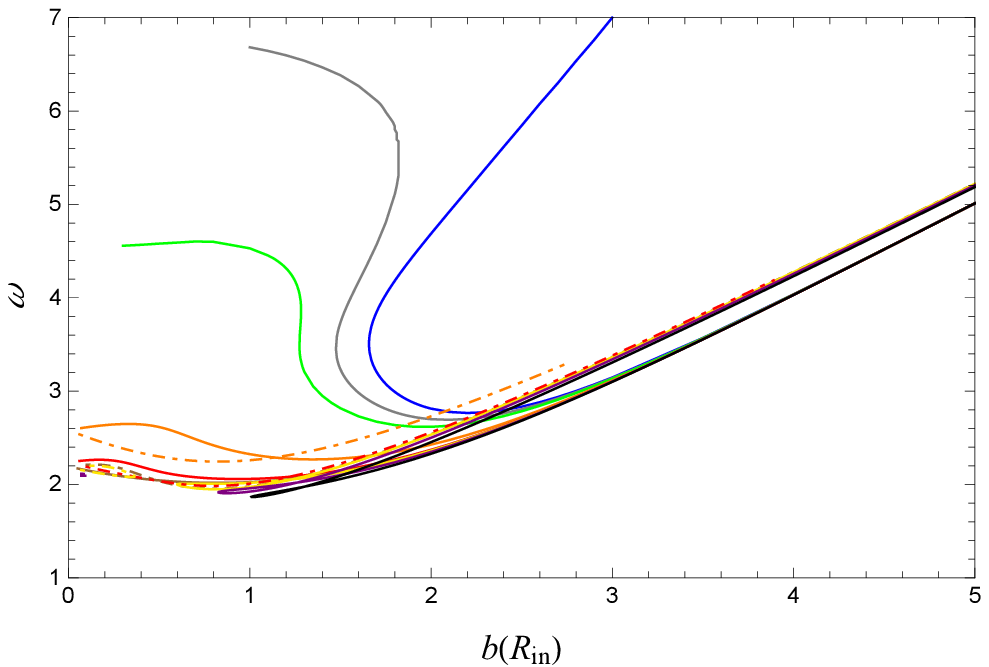}\\
\hspace{0.7cm}(a)\hspace{7.1cm}(b)\\
    \vspace{3mm}

     \includegraphics[width=80mm]{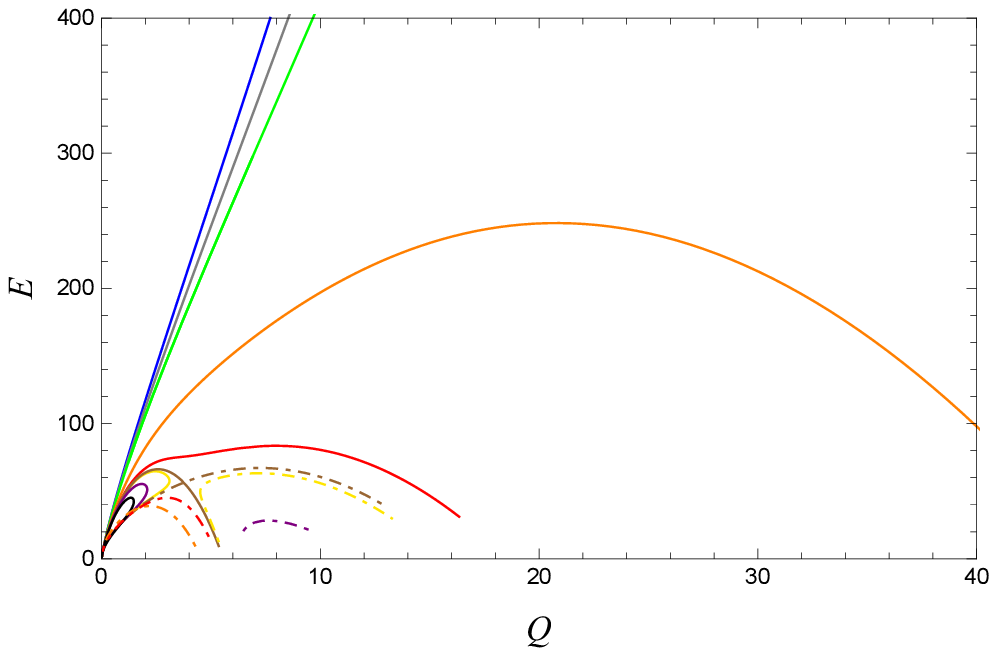}~~
     \includegraphics[width=80mm]{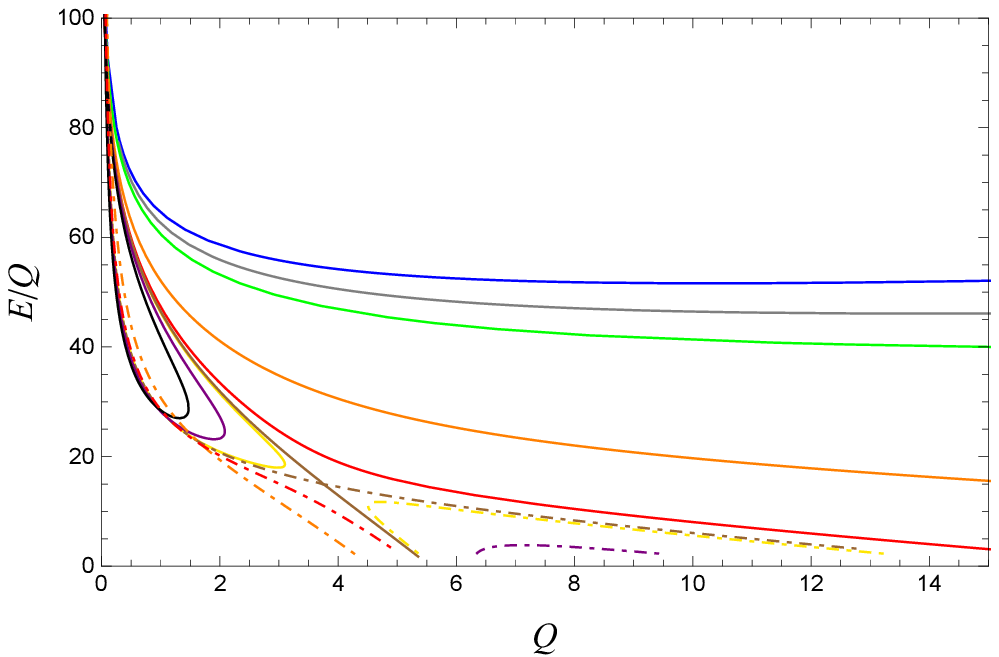}\\
\hspace{0.7cm}(c)\hspace{7.1cm}(d) \\
     \vspace{3mm}

    \caption{\label{CP11}   The $\mathbb{C}P^{11}$ boson shells.  
(a): The phase diagram of the ratio of inner and outer shell radii $R_{\rm in}/R_{\rm out}$ and 
value of the gauge field at the inner radius $b(R_{\rm in})$.
(c):~The relation between $E$ and $Q$. (d):~The relation between $E/Q$ and $Q$. 
}
\vspace{1cm}

\includegraphics[width=80mm]{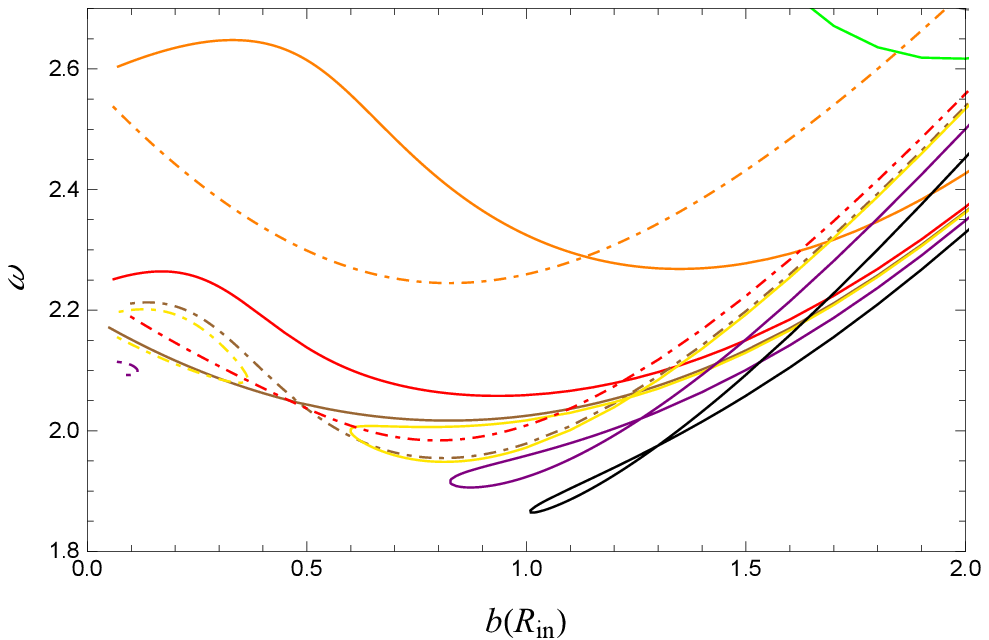}~~
\includegraphics[width=80mm]{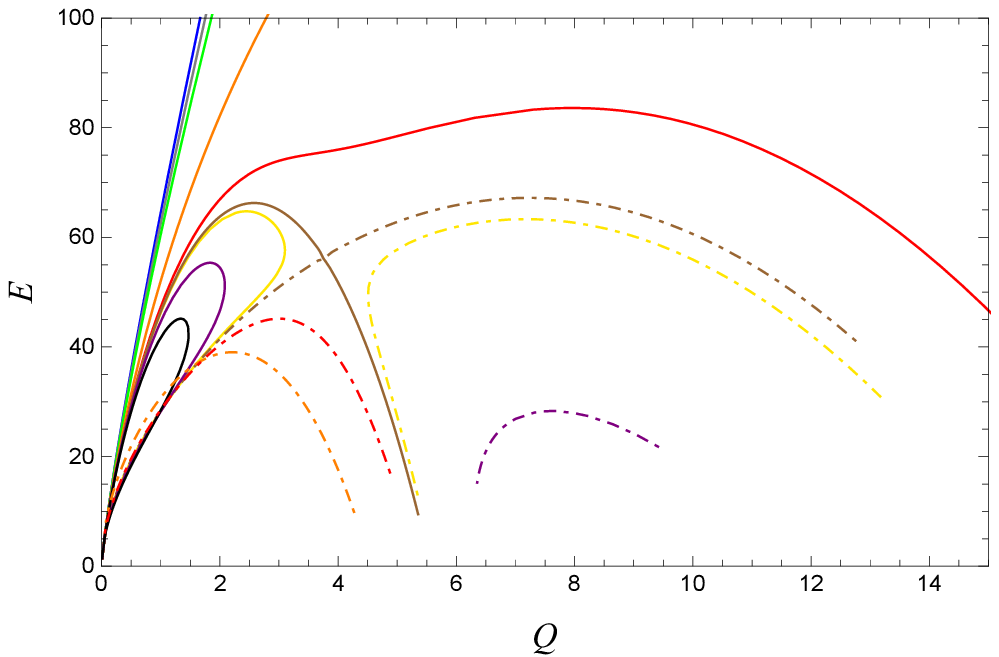}\\
\hspace{0.7cm}(a)\hspace{7.1cm}(b) \\
    \vspace{3mm}

\caption{\label{CP11large}   The  $\mathbb{C}P^{11}$ boson shells.  
(a):~The relation between the $b(0)$ and frequency $\omega$. The same as Fig.\ref{CP11}(b) but 
the plots at $0.0 \le b(R_{\rm in})\le 2.0,~1.8\le \omega \le 2.7$ are enlarged.
(b):~The relation between $E$ and $Q$. The same as Fig.\ref{CP11}(c) 
but the plots at $0.0 \le Q\le 15, 0\le E \le 100$ are enlarged. 
The  character and color of curves in this plot are consistent with Fig.\ref{CP11}(a). 
}

\end{center}
\end{figure*}


	\begin{figure*}[hbtp]
	\begin{center}

    \includegraphics[width=75mm]{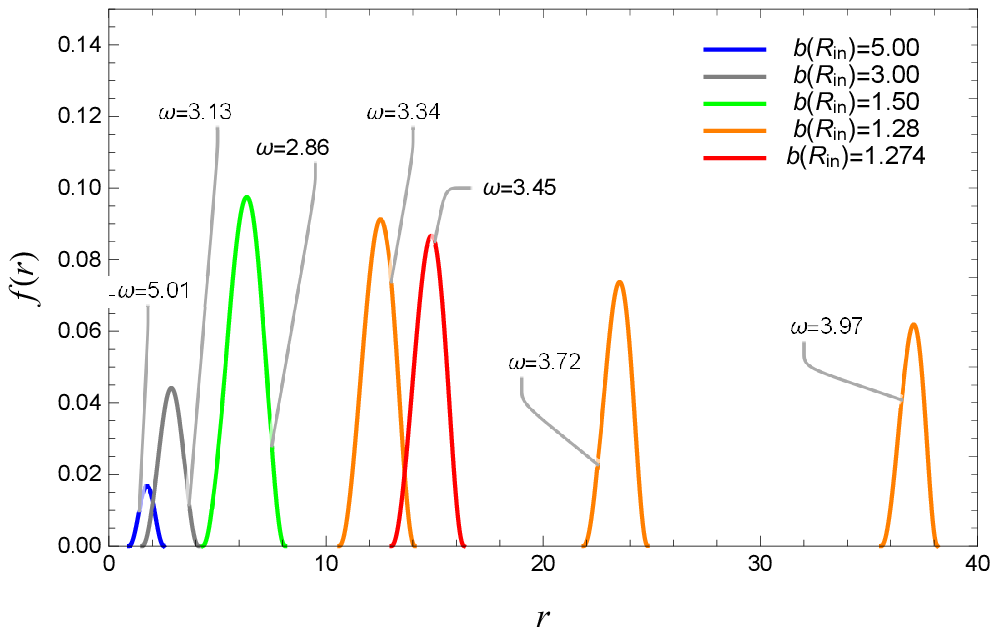}~~
    \includegraphics[width=75mm]{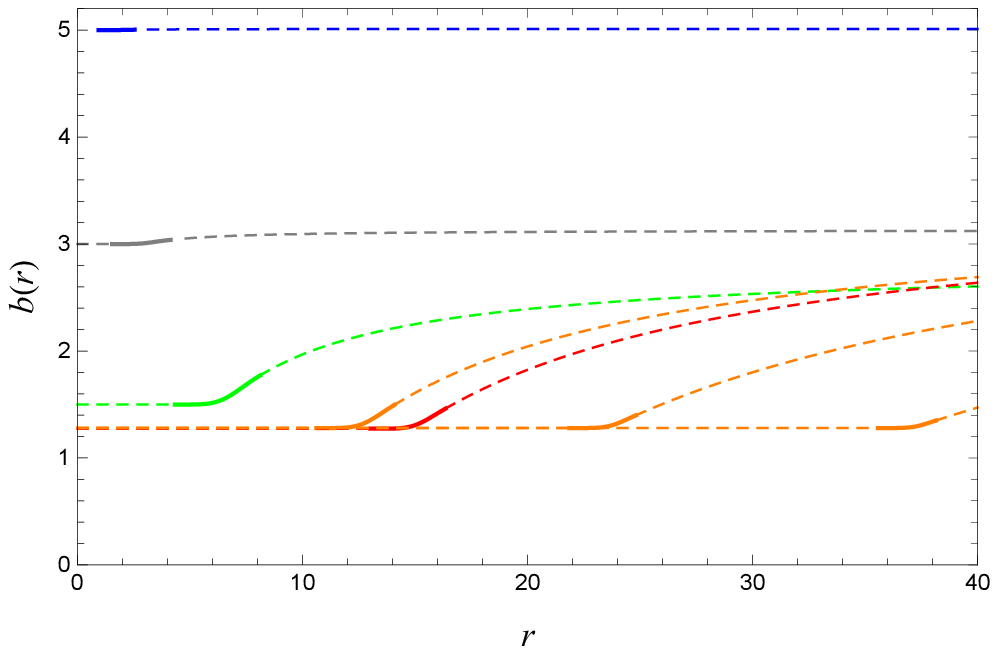}\\
\hspace{0.9cm}(a)\hspace{6.9cm}(b) 
\vspace{5mm}

    \includegraphics[width=75mm]{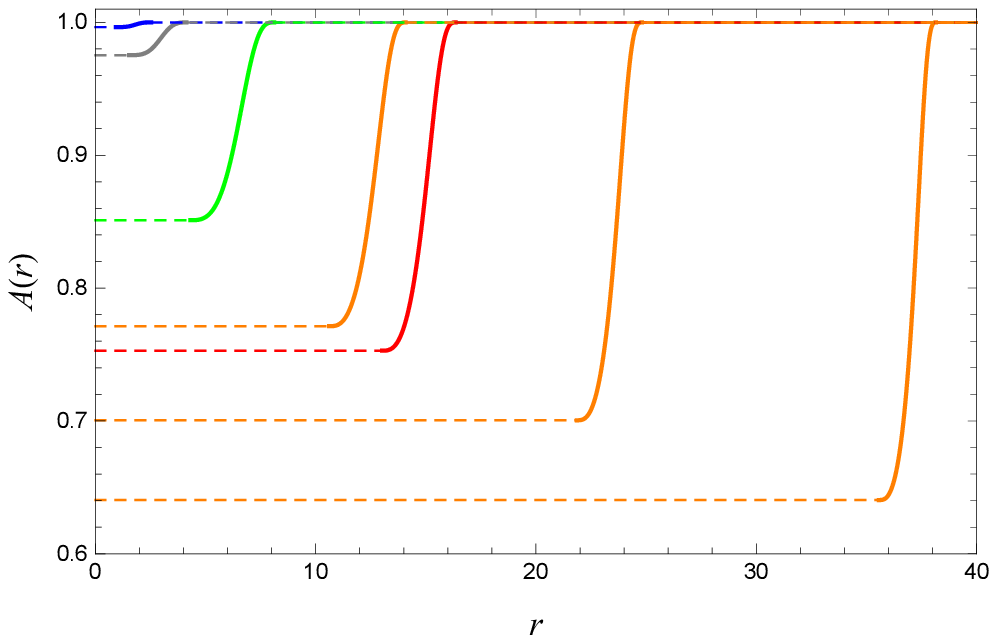}~~
    \includegraphics[width=75mm]{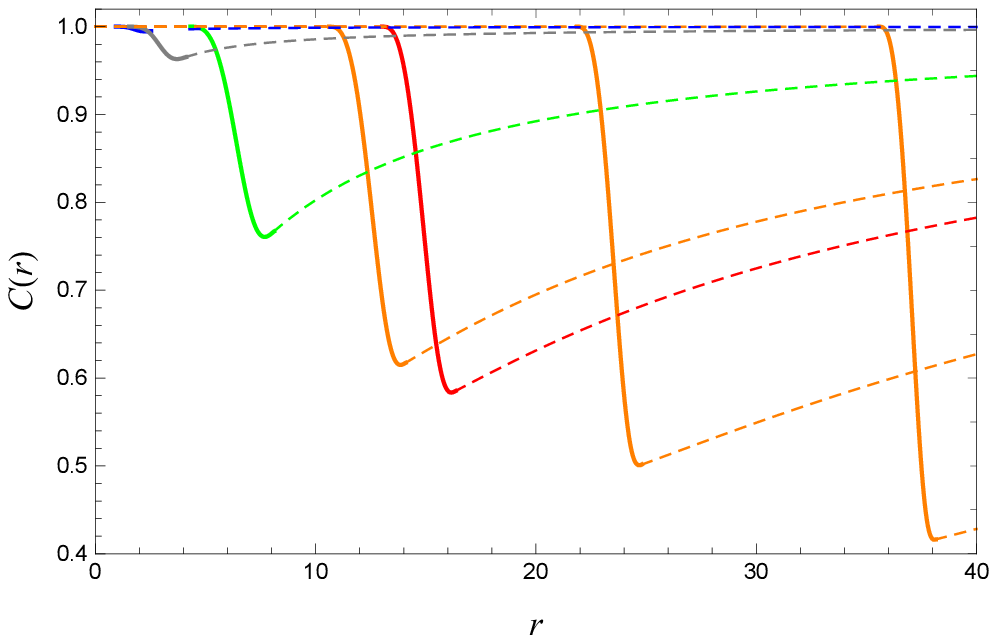}\\
\hspace{0.9cm}(c)\hspace{6.9cm}(d) 

    \vspace{0mm}

    \caption{\label{CP11fieldI01}The $\mathbb{C}P^{11}$ boson shell solutions from the region I, $\alpha=0.1$. 
(a): The scalar profile $f(r)$. 
(b): The gauge field $b(r)$. 
(c): The metric function $A(r)$. 
(d): The metric function $C(r)$. 
Solutions of the first branch are plotted with bold lines and the second branch are plotted with dot-dashed lines. 
Solutions of the equations where the scalar fields take the vacuum value are depicted with the dashed lines. }

\end{center}
\end{figure*}


	\begin{figure*}
\begin{center}

    \includegraphics[width=75mm]{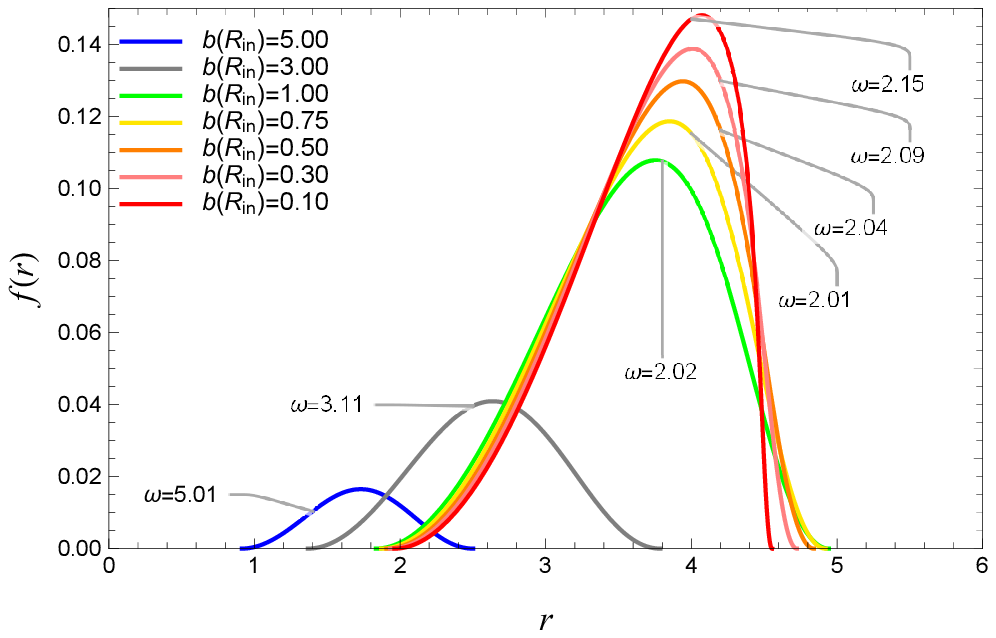}~~
    \includegraphics[width=75mm]{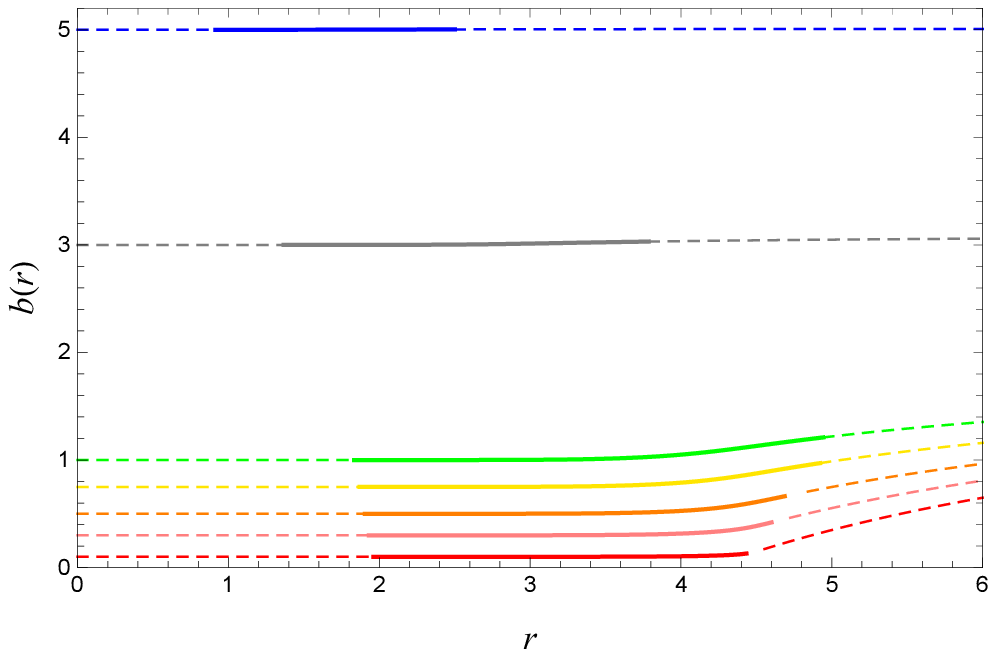}\\
\hspace{0.9cm}(a)\hspace{6.9cm}(b) 
\vspace{5mm}

    \includegraphics[width=75mm]{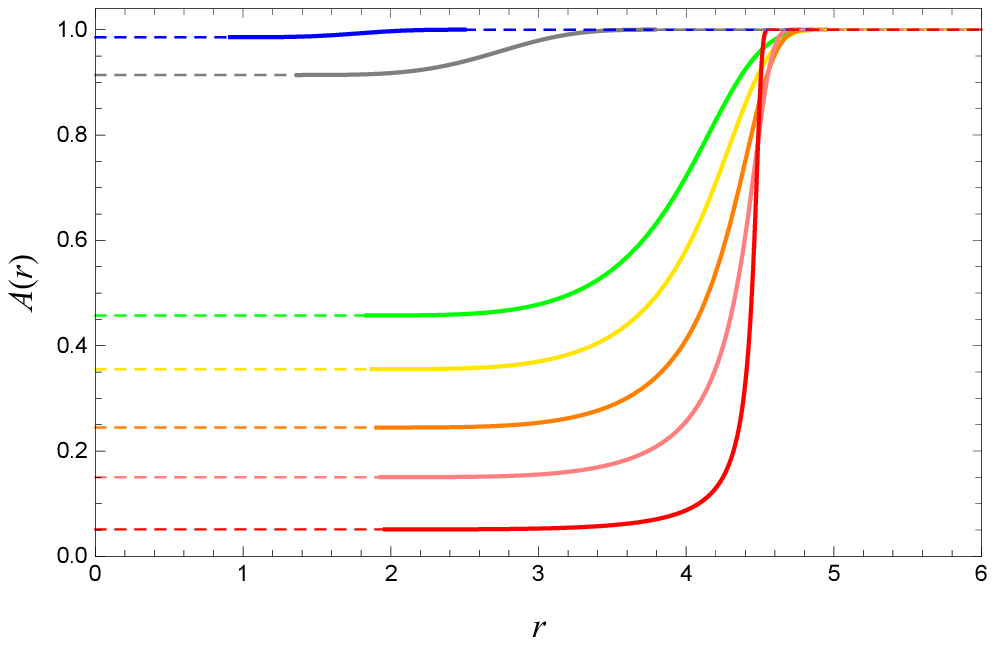}~~
    \includegraphics[width=75mm]{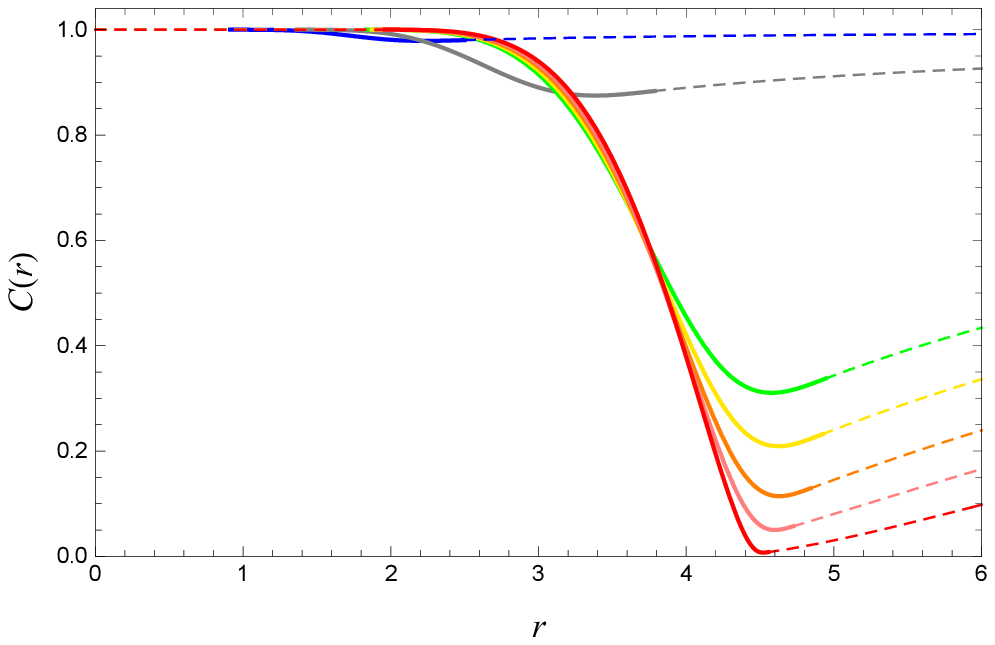}\\
\hspace{0.9cm}(c)\hspace{6.9cm}(d) 
    \vspace{0mm}

    \caption{\label{CP11fieldI04} The $\mathbb{C}P^{11}$ boson shell solutions from the region I, $\alpha=0.41625$. 
(a): The scalar profile $f(r)$. 
(b): The gauge field $b(r)$. 
(c): The metric function $A(r)$. 
(d): The metric function $C(r)$. 
Solutions of the first branch are plotted with bold lines and the second branch are plotted with dot-dashed lines. 
Solutions of the equations where the scalar fields take the vacuum value are depicted with dashed lines. }

\vspace{1cm}

    \includegraphics[width=75mm]{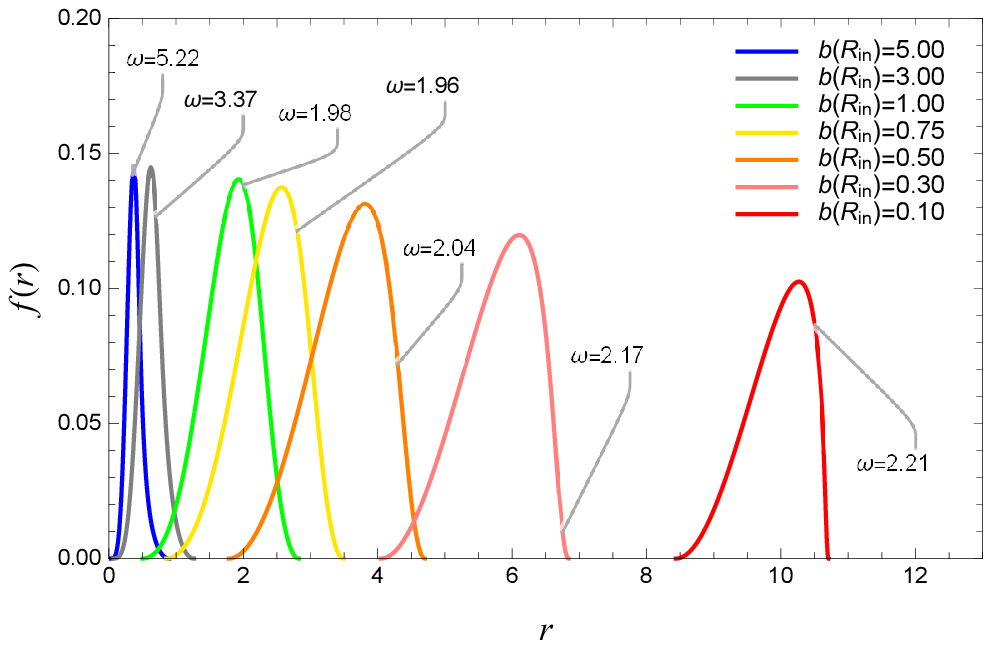}~~
    \includegraphics[width=75mm]{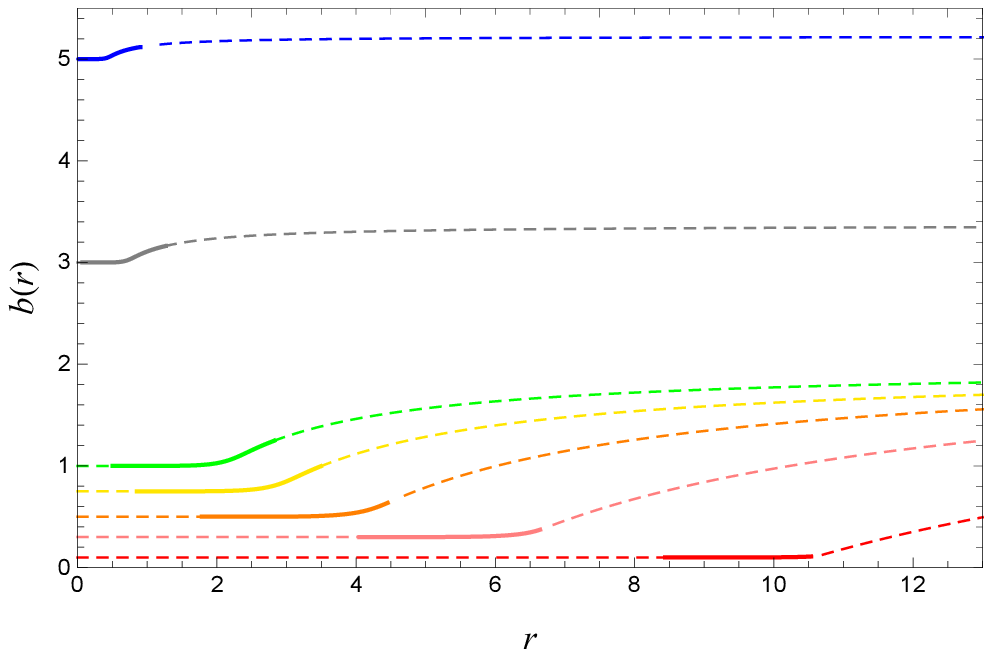}\\
\hspace{0.9cm}(a)\hspace{6.9cm}(b) 
\vspace{5mm}

    \includegraphics[width=75mm]{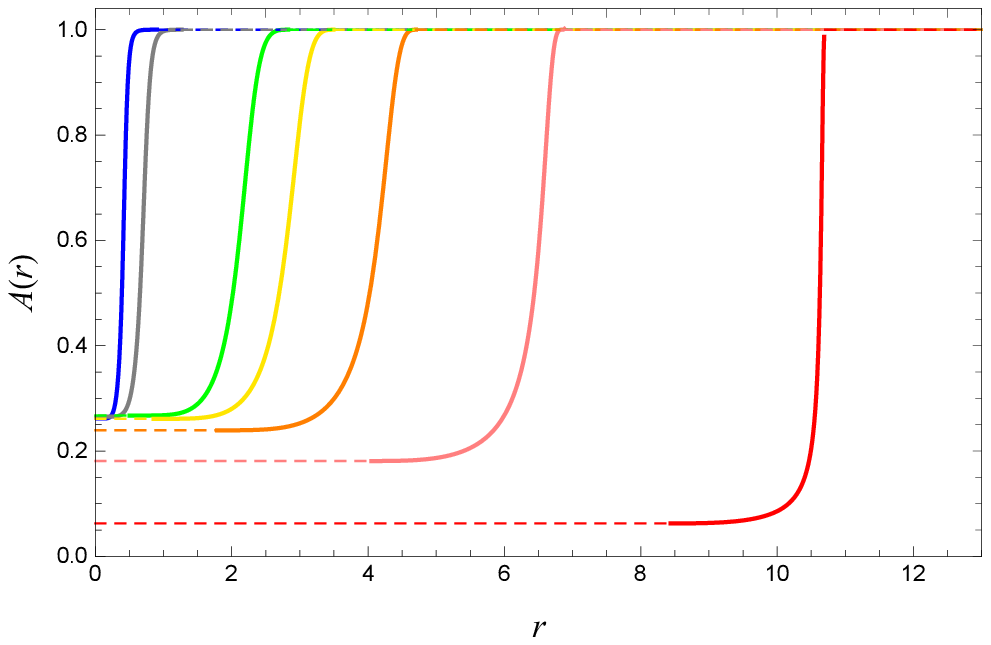}~~
    \includegraphics[width=75mm]{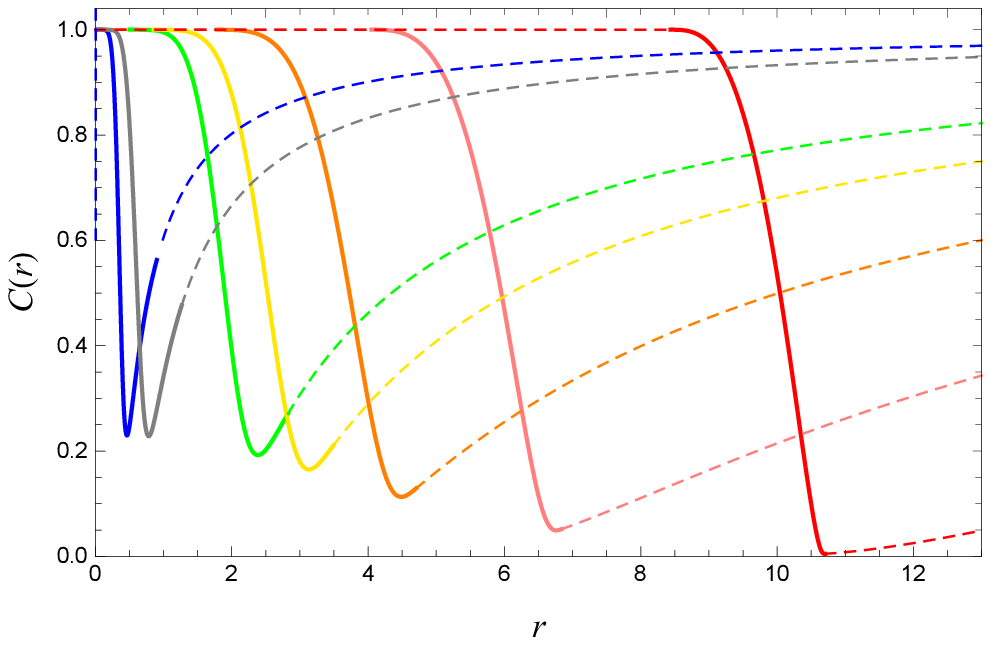}\\
\hspace{0.9cm}(c)\hspace{6.9cm}(d) 
    \vspace{0mm}

    \caption{\label{CP11fieldIa04} The $\mathbb{C}P^{11}$ boson shell solutions of the region Ia, $\alpha=0.41625$. 
(a): The scalar profile $f(r)$. 
(b): The gauge field $b(r)$. 
(c): The metric function $A(r)$. 
(d): The metric function $C(r)$. 
Solutions of the first branch are plotted with bold lines and the second branch are plotted with dot-dashed lines. 
Solutions of the equations where the scalar fields take the vacuum value are depicted with the dashed lines. }
  \end{center}
	\end{figure*}


\begin{figure*}[hbtp]
  \begin{center}
    \includegraphics[width=75mm]{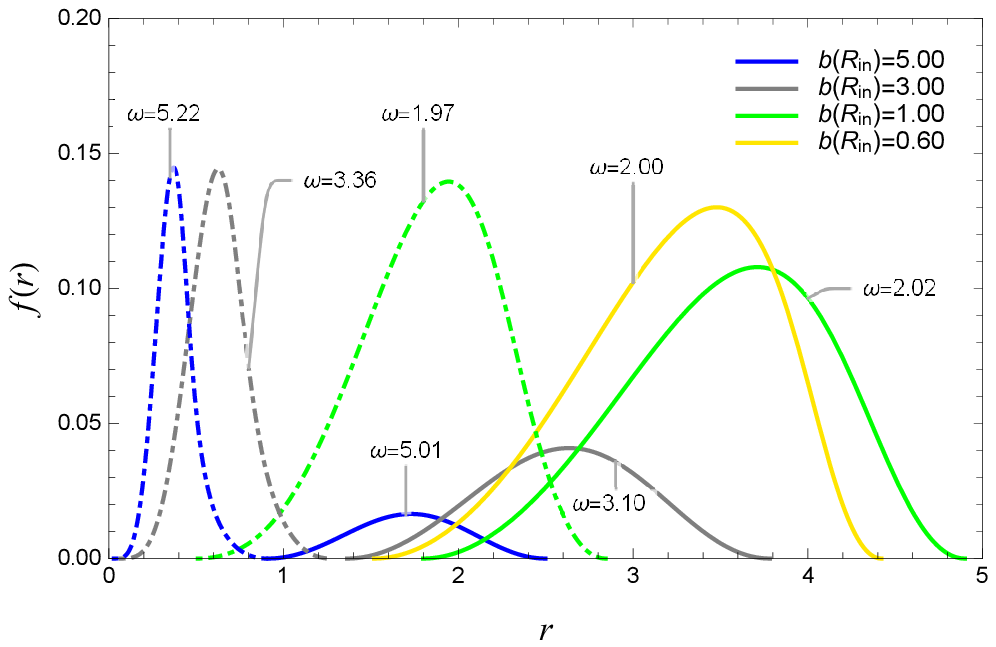}~~
    \includegraphics[width=75mm]{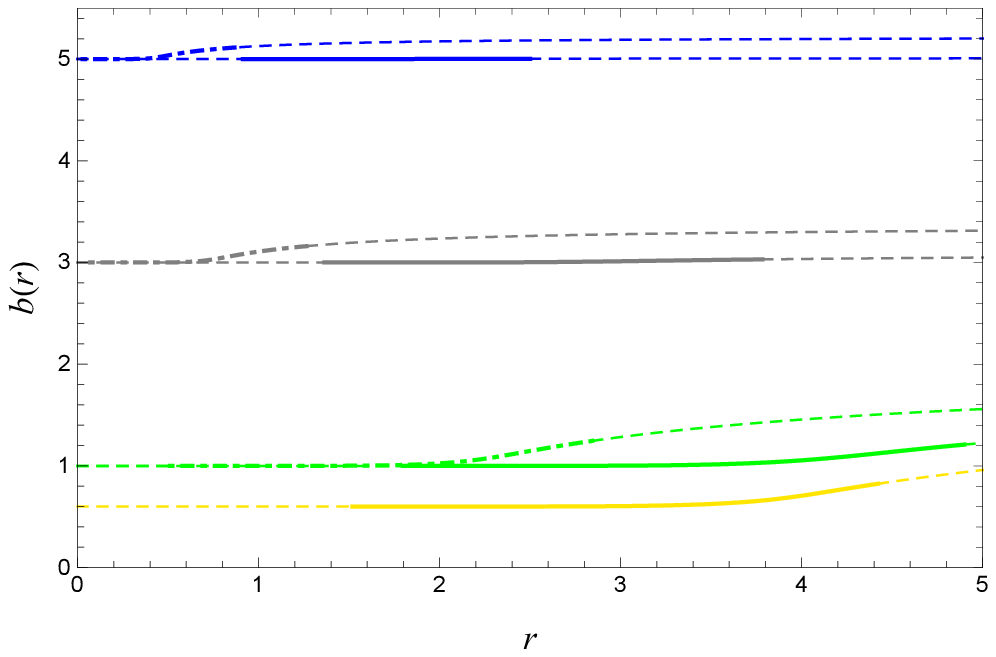}\\
\hspace{0.9cm}(a)\hspace{6.9cm}(b) 
\vspace{5mm}

    \includegraphics[width=75mm]{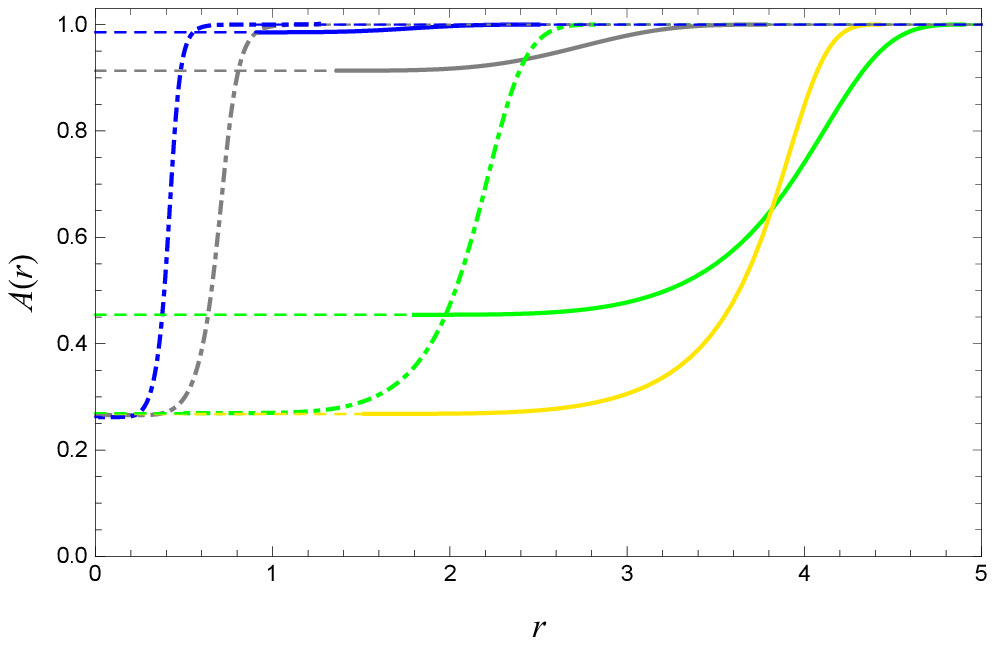}~~
    \includegraphics[width=75mm]{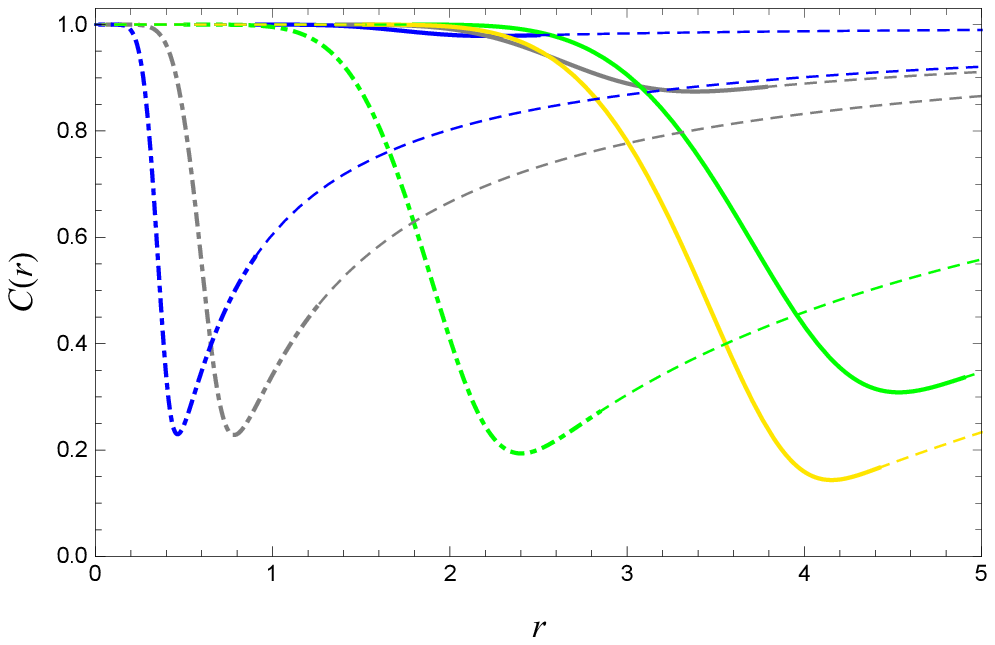}\\
\hspace{0.9cm}(c)\hspace{6.9cm}(d) 

    \vspace{0mm}

    \caption{\label{CP11fieldII042} The $\mathbb{C}P^{11}$ boson shell solutions from the region II, $\alpha=0.42$. 
(a): The scalar profile $f(r)$. 
(b): The gauge field $b(r)$. 
(c): The metric function $A(r)$. 
(d): The metric function $C(r)$. 
Solutions of the first branch are plotted with bold lines and the second branch are plotted with dot-dashed lines. 
Solutions of the equations where the scalar fields take the vacuum value are depicted with the dashed lines. }

\vspace{1cm}

    \includegraphics[width=75mm]{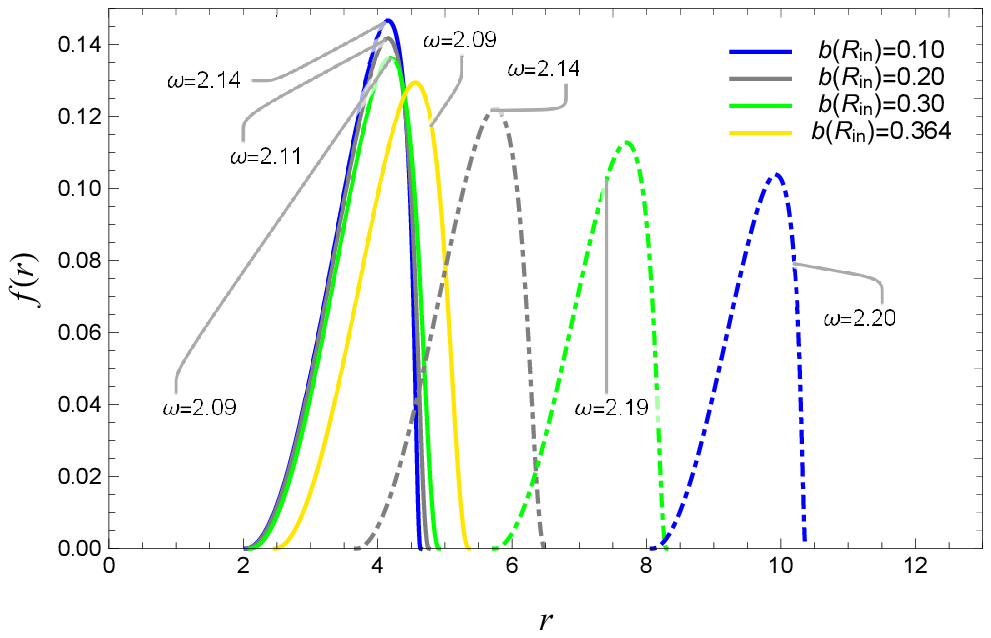}~~
    \includegraphics[width=75mm]{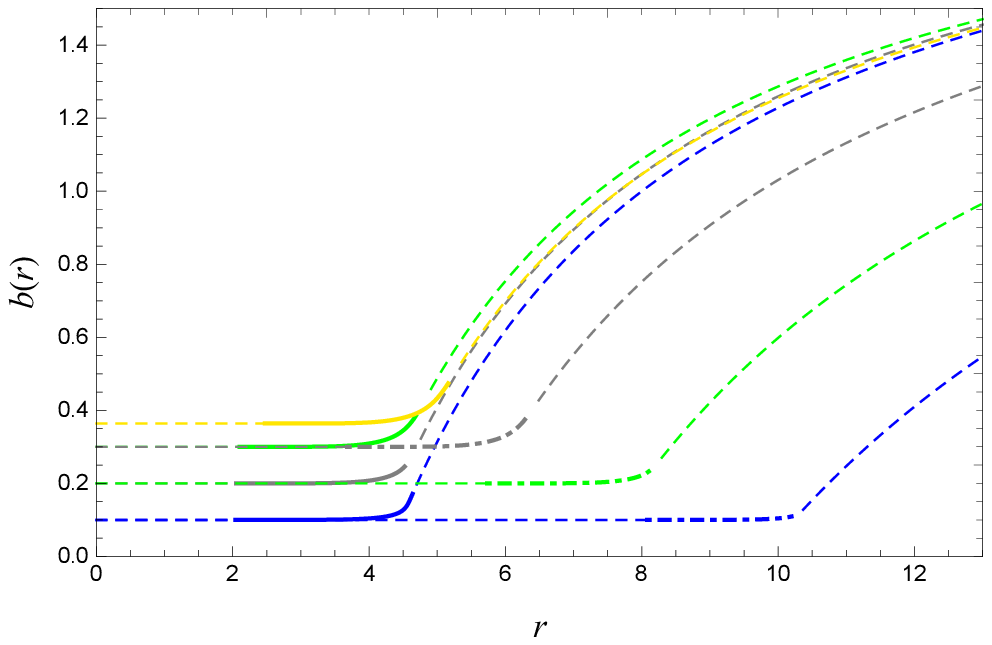}\\
\hspace{0.9cm}(a)\hspace{6.9cm}(b) 
\vspace{5mm}

    \includegraphics[width=75mm]{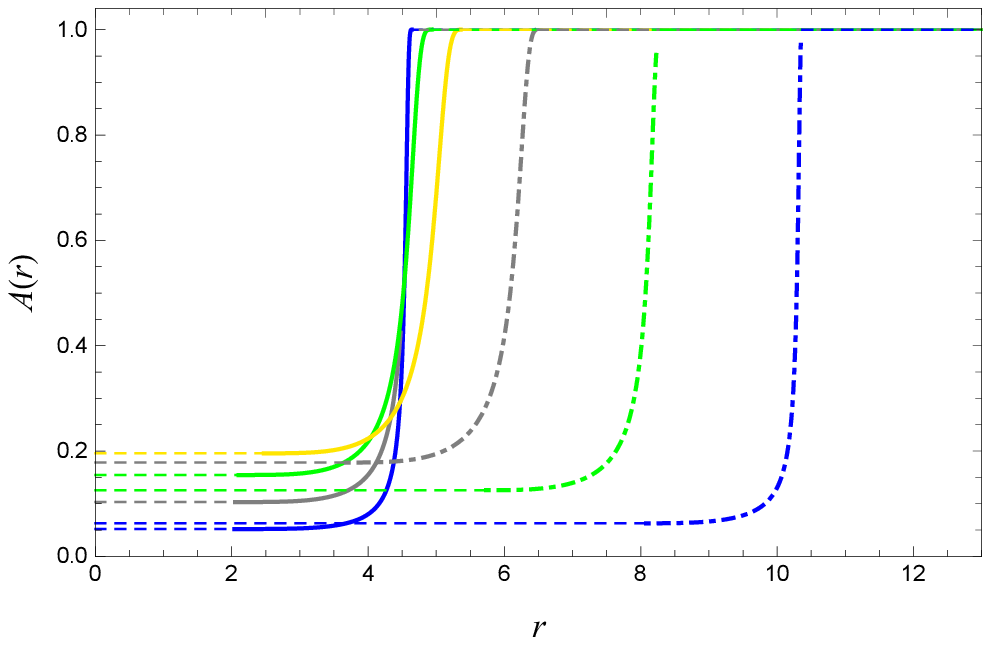}~~
    \includegraphics[width=75mm]{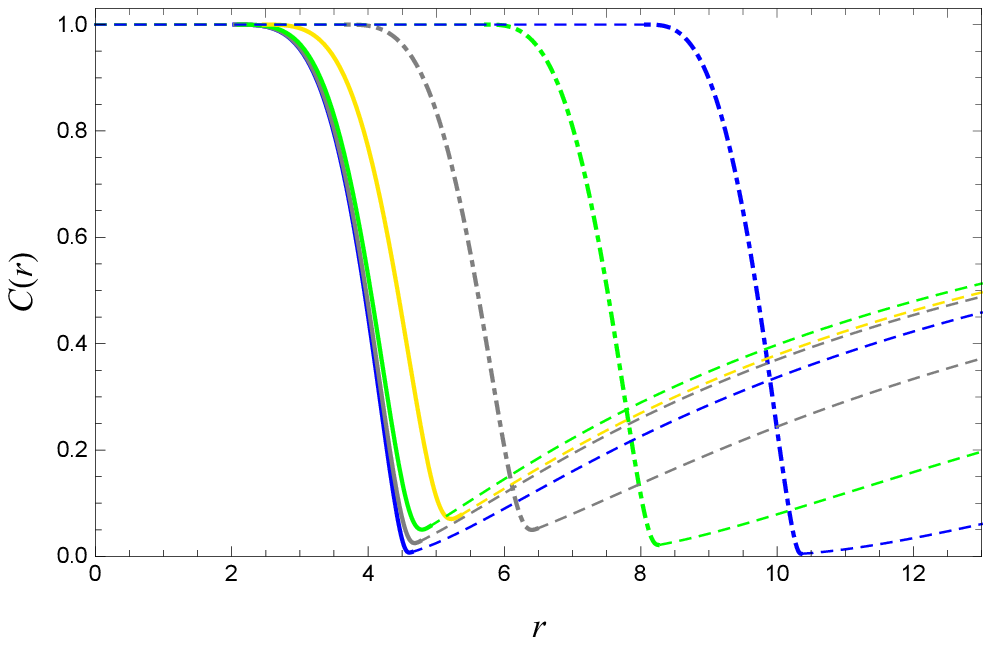}\\
\hspace{0.9cm}(c)\hspace{6.9cm}(d) 
    \vspace{0mm}

    \caption{\label{CP11fieldIIa042}~The $\mathbb{C}P^{11}$ boson shell solutions of the region IIa, $\alpha=0.42$. 
(a): The scalar profile $f(r)$. 
(b): The gauge field $b(r)$. 
(c): The metric function $A(r)$. 
(d): The metric function $C(r)$. 
Solutions of the first branch are plotted with bold lines and the second branch are plotted with dot-dashed lines. 
Solutions of the equations where the scalar fields take the vacuum value are depicted with the dashed lines. }
  \end{center}
	\end{figure*}


	\begin{figure*}[t]
	\begin{center}

    \includegraphics[width=80mm]{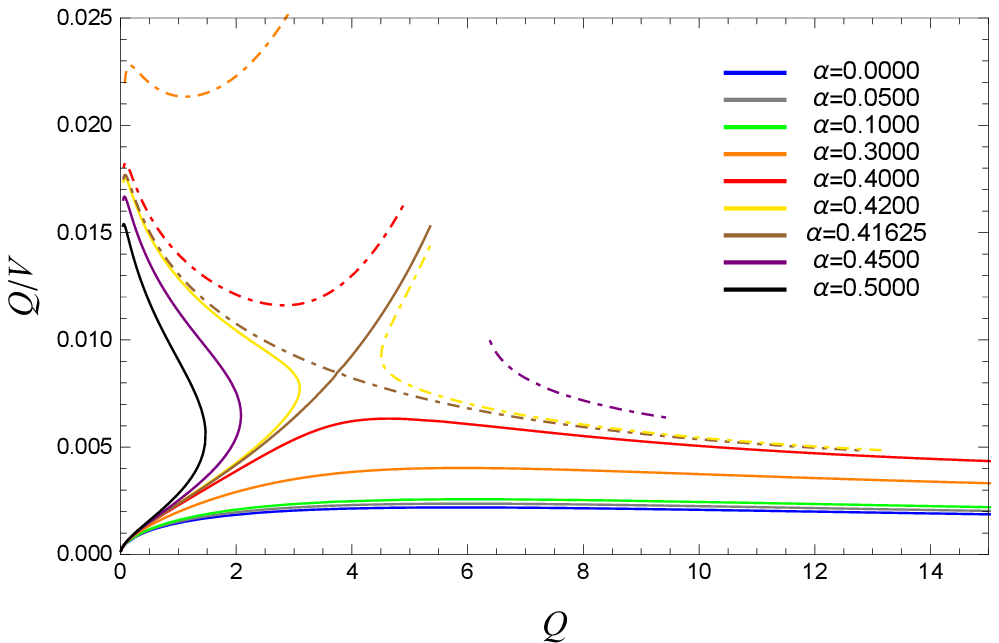}~~
\includegraphics[width=80mm]{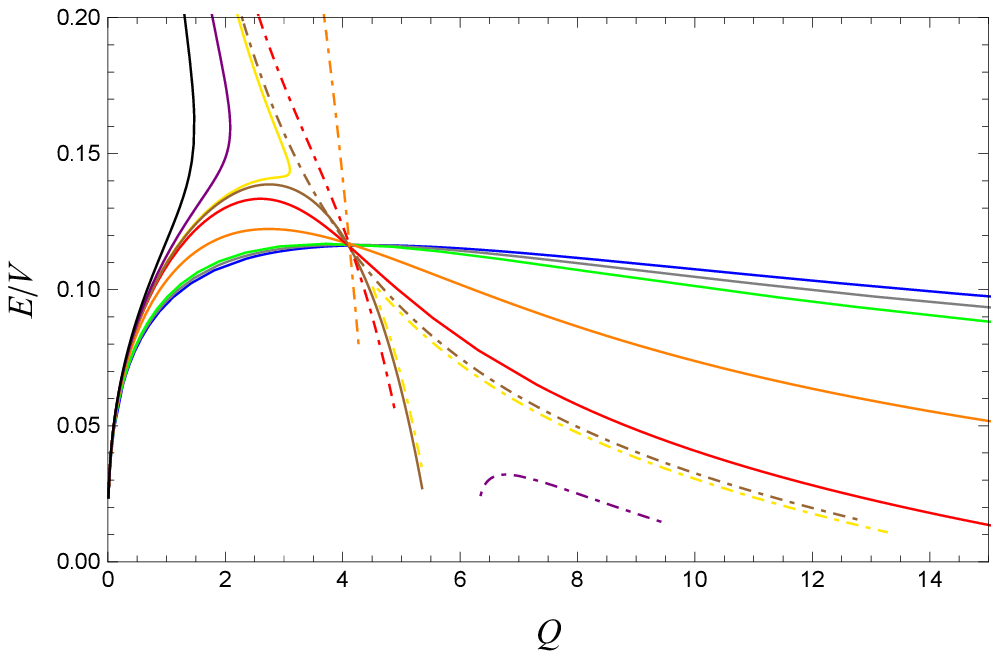}
\\
\hspace{0.9cm}(a)\hspace{7.1cm}(b) 
    \vspace{5mm}

\includegraphics[width=80mm]{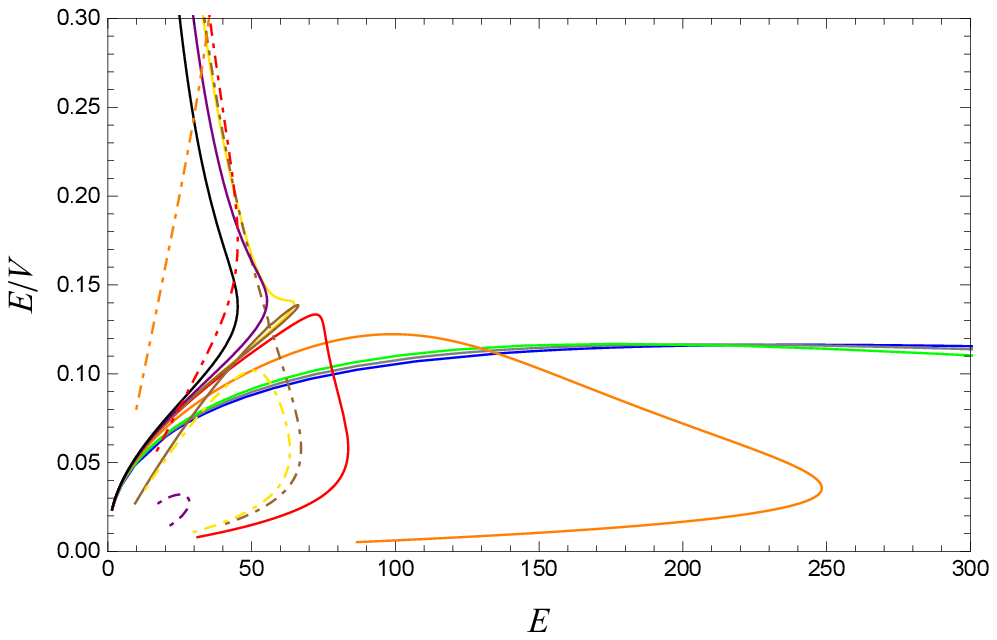}\\
\hspace{0.9cm}(c) 
    \vspace{3mm}

    \caption{\label{CP11Qd} The $\mathbb{C}P^{11}$ boson shell solutions. 
(a):~The charge density: (the Noether charge $Q$)/(the volume of the $Q$-shell $V$) and the charge $Q$.
(b):~The energy density: (the total energy of $Q$-shell)/(the volume of the $Q$-shell $V$) and the charge $Q$.
(c):~The energy density and the energy $E$. }
  \end{center}
	\end{figure*}


	\section{The $\mathbb{C}P^1$ ball and shell}
	
	We first study the case of $n=0~(N=1)$, in which the globally regular $Q$-ball and  also the shell emerge. 
	Figure \ref{CP1phase}(a) represents the phase diagram of the $Q$-ball for the values of the fields, 
	{\it i.e.}, the scalar profile $f(0)$ and the gauge function $b(0)$ by changing the gravitating coupling constant $\alpha$. 
	The behavior shares the basic feature with results of a single complex scalar field model
	~\cite{Kleihaus:2009kr,Kleihaus:2010ep,Kumar:2014kna,Kumar:2015sia,Kumar:2016sxx}.
	There appear four regions of the solutions and according to the previous studies, we call them as I,~Ia,~II, and IIa.  
	Here we show the results for our $\mathbb{C}P^1$ model.  
	For $\alpha=0$, the $Q$-ball solutions are represented by a blue line in the lower part of the figure.The 
	solutions are characterized via a maximum of $f(0)$ (dots) and a minimum of $b(0)$ (triangles). 
	After the minimum, for increasing the frequency $\omega$, the solutions move to $f(0)=0$ (the crosses), 
	where the solutions are maximally delocalized from the origin. 
	Here, the points (the crosses) are the bifurcation points with shell-like solutions.

	When we switch on the gravitating coupling constant $\alpha$, the solutions tend to move inside, where the maximum of $f(0)$ grows 
	and the minimum of $b(0)$ reduces. Just above a critical value $\alpha\sim 0.7$, 
	the solutions do not reach $f(0)=0$ any longer but take some finite values at the center. 
	Apparently, it reflects the fact that the gravitational force has an attractive nature. 
	All these solutions form region I. 
	
	As the further evolution continues, the solutions reach to the next bifurcation with a second set of solutions, the boundary of region Ia, 
	at the critical value $\alpha_{\rm crit}=0.8094$. The solutions of Ia exist for the coupling constant $\alpha\le \alpha_{\rm crit}$
	and the minimum of $f(0)$ (the diamonds) increases with decreasing $\alpha$. 
	
	At $\alpha_{\rm crit}$ the solutions of I and Ia bifurcate and,  for $\alpha>\alpha_{\rm crit}$ they split into a right II 
	and left IIa regions. The solutions of II correspond to larger $b(0)$ and that of IIa to smaller $b(0)$. With increasing $\alpha$, 
	the solutions of IIa
	move to smaller values of $b(0)$ and disappear at some critical $\alpha$. The solution of II move toward larger $b(0)$. 

	After passing the bifurcation points (the crosses), the shell-like solutions emerge. Inside the hollow region of the shell
 	$0\le r < R_{\rm in}$, 
	the gauge field $b(r)$ is constant and the profile $f(r)$ vanishes. Correspondingly, the space-time is Minkowski-like, i.e., 
	$A(r)=$const., $C(r)=1$. in $0\le r<R_{\rm in}$. 
	Outside the shell $R_{\rm out}$, the space-time becomes a Reissner-Nordstr\"om.  
	Figure \ref{CP1phase}(b) is the plot of the relation between $E$ and $Q$ and  
	Fig.\ref{CP1phase}(c) is the relation between $b(0)$ and the frequency $\omega$. 
	We present the enlarged plot of (c) in Fig.\ref{CP1phase}(d).
	Figure \ref{CP1phase}(e) presents the phase diagram of the shell-like solution: 
	the ratio of inner and outer shell radii $R_{\rm in}/R_{\rm out}$ as value of the gauge field at the inner radius $b(R_{\rm in})$. 
	As $\omega$ increases, the solutions delocalize from the origin and 
	also $b(R_{\rm in})$ grows. For finite gravitational coupling constant, a throat is formed at the outer radius $R_{\rm out}$ 
	and the value of $b(R_{\rm in})=b(0)$ reaches zero. In Fig.\ref{CP1phase}(f), we plot the relation between $E$ and $Q$ for 
	both the shell (the bold line) and the ball (the dashed line) in region I.

	\begin{figure*}[htbp]
	\begin{center}

    \includegraphics[width=80mm]{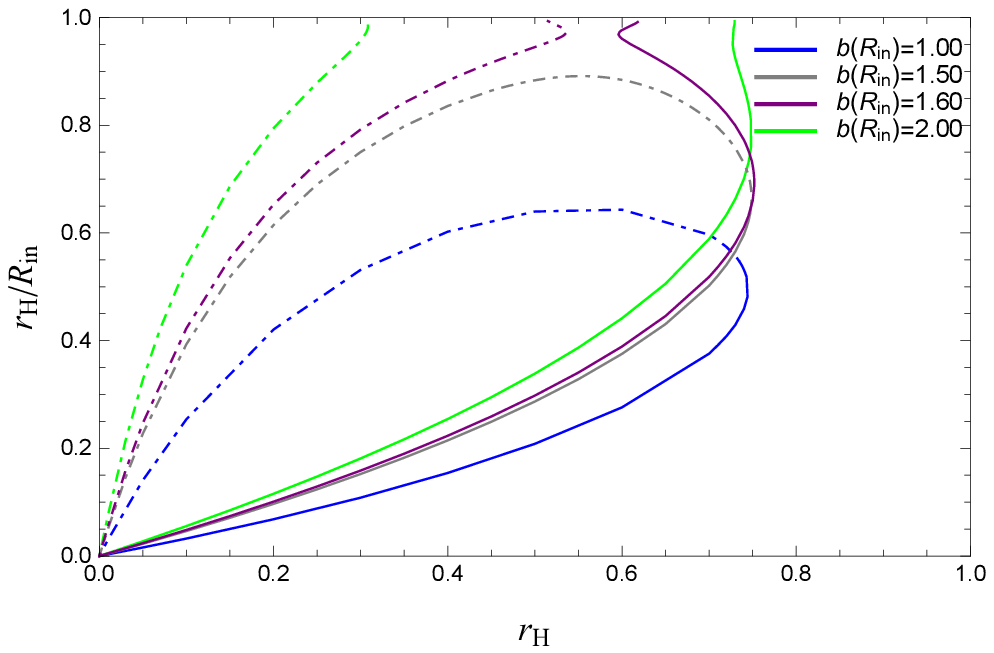}~~
\includegraphics[width=80mm]{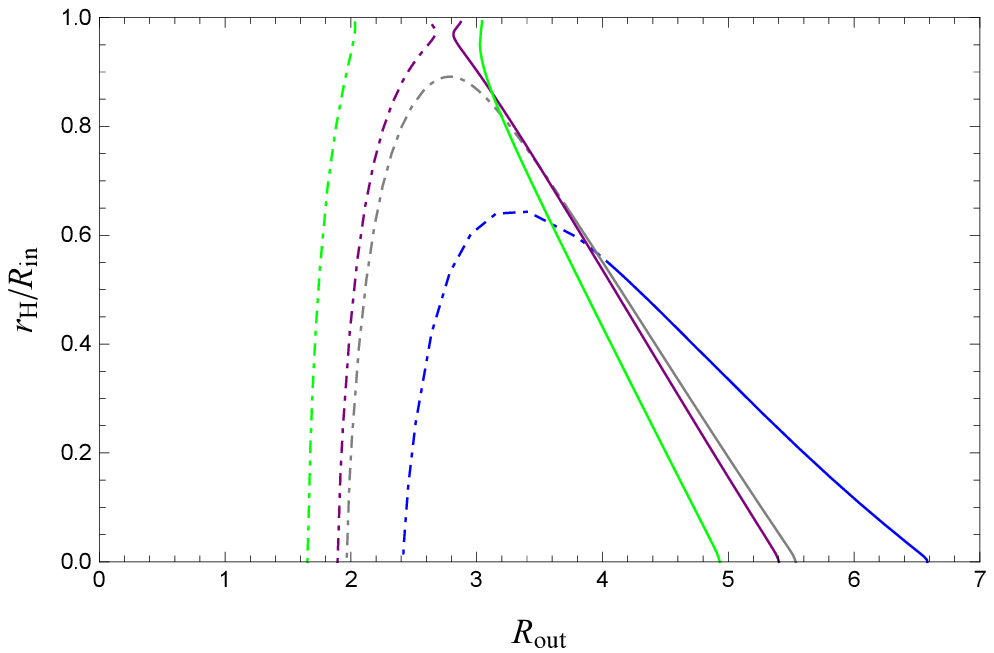}
\\
\hspace{0.5cm}(a)\hspace{7.2cm}(b) 

    \caption{\label{Harbor}~The $\mathbb{C}P^{11}$ harbor solution with the black hole with the charge $Q_{\rm H}=0.001, \alpha=0.30$. 
(a):~The ratio (the horizon radius)/(the inner radius of the shell) $r_{\rm H}/R_{\rm in}$ and the $r_{\rm H}$. 
(b):~The $r_{\rm H}/R_{\rm in}$ and the outer radius $R_{\rm out}$. The solutions of the region I are plotted with bold lines and 
the Ia are the dot-dashed lines. 
 }
  \end{center}
	\end{figure*}


	\section{The boson shells}
	The solution of $n=1$ has almost similar properties with $n=0$. Therefore, 
	we next consider the boson shell configurations in $n\ge 2$. 
	We show the case of $n=5~(N=11)$. In Fig.\ref{CP11}(a), we plot the phase diagram of
	the ratio of inner and outer shell radii $R_{\rm in}/R_{\rm out}$ as 
	value of the gauge field at the inner radius $b(R_{\rm in})$. There are four distinct regions labeled by
	I,~Ia,~II,~and IIa. 
	Figure \ref{CP11}(b) presents the relation between the frequency $\omega$ and $b(R_{\rm in})$.
	Figure \ref{CP11}(c) is the relation between the energy $E$ versus the charge $Q$. 
	Also, we plot the ratio $E/Q$ versus the charge $Q$, which is shown in Fig.\ref{CP11}(d).
	These are useful to demonstrate the notable difference of the behavior of the weak gravity (small $\alpha$) and 
	the strong gravity (large $\alpha$). For small $\alpha$, 
	the energy $E$ monotonically increases with $Q$ and the ratio $E/Q$ monotonically decreases with $Q$.
	For large $\alpha$, some remarkable differences appear on the plots. 
	In Fig.\ref{CP11large}, we show the enlarged plots of (b) and (c) of Fig.\ref{CP11}. 
	The solutions form closed loops or knots for $\alpha>\alpha_{\rm crit}$. 

	Let us discuss the solutions of each region in detail. 		
	We begin with the solutions of region I. 
	In Figs.\ref{CP11fieldI01} and \ref{CP11fieldI04}, we plot the region I solutions of the 
	scalar profile $f(r)$ and the gauge field $b(r)$ and also the metric functions $A(r)$ and $C(r)$ 
	for $\alpha=0.1$ and $\alpha=0.41625$.   
	Figure \ref{CP11fieldI01} indicates that for smaller $\omega$ the solutions of the first branch move outward 
	and join the second branch {\it i.e.}, for larger $\omega$
	the solutions accelerate to go outward with decreasing $b(R_{\rm in})$.  
	Finally the ratio $R_{\rm in}/R_{\rm out}$ achieves 
	unity because the size of the solutions rapidly expands with keeping the thickness. 
	Similar to the case of $n=0$, the throat forms and the value of $b(R_{\rm in})$ goes to zero.  
	However, after a critical point of the coupling constant, the ratio $R_{\rm in}/R_{\rm out}$ remains 
	some fractional values.  
	In Fig.\ref{CP11fieldI04}, we show a peculiar example, where the coupling constant is a critical value
	$\alpha_{\rm crit}=0.41625$. For large $\omega$ [or $b(R_{\rm in}$)], the solutions move outward for reducing $b(R_{\rm in})$ 
	but after a critical point $b(R_{\rm in})=1.0$, the solutions almost are stuck and
	the ratios hold a definite value. 
	
	As the further evolution continues, the solutions reach a bifurcation with a second set of solutions, 
	the boundary of region Ia, 
	at the critical value $\alpha_{\rm crit}=0.41625$. The solutions of Ia exist for the coupling constant 
	$\alpha\le \alpha_{\rm crit}$ and are the shell with finite thickness. The solutions are shown in Fig.\ref{CP11fieldIa04}. 
	
	At $\alpha_{\rm crit}$ the solutions of I and Ia bifurcate, and for $\alpha>\alpha_{\rm crit}$ they split into right II 
	and left IIa regions. They are novel solutions that exist in our $\mathbb{C}P^N$ model. 
	The solution of II for $\alpha=0.42$ is Fig.\ref{CP11fieldII042} and that of IIa is Fig.\ref{CP11fieldIIa042}. 
	On the first branch, the solutions of II move outward with reducing $b(R_{\rm in})$ for decreasing $\omega$ (the bold line) 
	and after passing a minimum of $b(R_{\rm in})=0.60$, 
	the second branch (dot-dashed line) inverts to go inside. 
	The solutions tend to shrink then and the ratio $R_{\rm in}/R_{\rm out}$ becomes smaller. 
	The behavior of the solutions of IIa is opposite. The first branch is almost stacked and the second branch moves outward.
	
	For larger $n~(>5)$, the behavior is almost similar. 
	The only difference is that in the phase diagram of $R_{\rm in}/R_{\rm out}$ versus 
	$b(R_{\rm in})$, especially the area of the region Ia grows as $n$ increases. 
	To be specific, the solutions of $\mathbb{C}P^{51}$ show that 
	the border line between the areas I,IIa and Ia,II 
	[in the case of $\mathbb{C}P^{11}$ the bold line of $\alpha=0.41625$ of Fig.\ref{CP11}(a)] 
	rises the position
	and the line of I,II and Ia, IIa (the dot-dashed line of $\alpha=0.41625$) 
	moves to the right.

	In summary, for a small coupling constant both the solutions of I (the ground state) 
	and Ia (the excited state) move outward as $\omega$ increases. 
	In the ground state, thickness of the shells goes to a definite constant by increasing the inner radius, and then they finally 
	become a shell with vanishing thickness. 
	On the other hand, for the excited state, the shell tends to become thicker. 	
	For a large coupling constant $\alpha>\alpha_{\rm crit}$, 
	the solutions shrink after passing a critical point. 
	The reason why such a complicated behavior and bifurcation are present is the interplay between 
	the electric force and the gravity.   
	We shall give a detailed discussion in the next section.

	\section{Further discussions}
	As it was seen in the previous section, our boson shells exhibit distinctive behavior depending on the value of the coupling constant $\alpha$. 	
	In order to see qualitatively the mechanism, we examine the energy density and the charge density of our solutions. 
	Thanks to the compactness of the solutions, we can directly compute 
	the volume of the $Q$-shells in terms of the compacton radius $r=R_{\rm in},R_{\rm out}$. 
	In Fig.\ref{CP11Qd}, we present the behavior of the charge density and the energy density versus the charge $Q$ or the energy $E$. 
	Figure \ref{CP11Qd}(a) shows the charge density versus the charge. 
	For region I solutions (small $\alpha$), the density is small and the change is moderate.
	The density grows as $\alpha$ increases which reflects the attractive nature of the gravity.
	 	
	When the solutions reach a bifurcation with region II, they exhibit a distinct characteristic feature; 
	now the density suddenly grows while the charge decreases, 
	which is originated from the fact that the solution quickly shrinks as $\alpha$ increases.
	The compactons become quite small objects $-$ the so-called mini $Q$-shells.
	In Fig.\ref{CP11Qd}(b), we show the energy density versus the charge, which looks similar to the charge 
	density case. 
	For the energy density as a function of the energy, the effect of the gravity is more apparent.
	Figure \ref{CP11Qd}(c) presents the energy density versus the energy. 
	For small coupling constants, the change of the density is more moderate. After some critical points, 
	the solution begins to fold, {\it i.e.}, both the energy and the energy density go to a smaller value.  
	The behavior of the critical solution ($\alpha_{\rm crit}=0.41625$) is particularly interesting; 
	it indicates that the volume is nearly constant which clearly shows a balance between the electric 
	force and the gravity. 
	Since the energy density tends to be a singular function for higher $\alpha_{\rm crit}$, it indicates 
	that the shell shrinks due to the dominating character of the gravity. 
	
      When we replace the inner empty Minkowski space of the shell by a charged black hole, the solutions become the
	harbor. 
	Some characteristics of the harbor solutions are presented in Fig.\ref{Harbor}. 
	In Fig.\ref{Harbor}(a), we show the ratio of the horizon radius $r_{\rm H}$ and the inner radius $R_{\rm in}$ as a function 
	of the $r_{\rm H}$
	and Fig.\ref{Harbor}(b) presents the same ratio but for the outer radius $R_{\rm out}$. 
	For finding these solutions, we compute the equations for the fixed black hole charge $Q_{\rm H}$ with changing $r_{\rm H}$.     
	As it was expected, the ratio $r_{\rm H}/R_{\rm in}$ is close to 1 but not exact, which means the solutions are just the harbor, 
	not the hair. 
	Though the solutions of I and Ia had been independent, now the black holes smoothly connect them. 
	For a small $b(R_{\rm in})$, {\it i.e.}, $b(R_{\rm in})=1.00, 1.50$, 
	the solutions I and Ia become continuous at some critical $r_{\rm H,crit}$.
	As a result, regions I, Ia are integrated by the harboring black holes. 	
	Above some critical point of $b(R_{\rm in})$, they separate off and never touch each other. 
	For these solutions, there are forbidden values of $r_{\rm H}$ where $r_{\rm H}/R_{\rm in}$ is 
	beyond the unity. 
	As a result, regions I, Ia become isolated in these cases. 
	A similar mechanism exists for region II where  
	two independent solutions with same $\alpha$ at some value  of $b(R_{\rm in})$ exist. 
	When we introduce an event horizon $r=\tilde{r}_{\rm H,crit}$, they are merged with each other. 
	Similarly, as above, for large $b(R_{\rm in})$ they remain isolated. 
		
	\section{Conclusions}

We presented the several phase diagrams for the $U(1)$ gauged $\mathbb{C}P^N$ nonlinear sigma model 
coupled with gravity. 
We obtained the compact $Q$-ball and $Q$-shell solutions in the standard shooting method. 
The resulting self-gravitating regular solutions form boson stars and boson shells. 
For the compact $Q$-shell solutions we put black holes in the interior 
and the exterior of the shell became the Reissner-Nordstr\"om space-time, which is called the harbor of the black holes.  
For several quantities of the solutions, characteristic phase diagrams were investigated. 
We observed four distinct regions, {\it i.e.}, regions I, Ia, II, and IIa. 
For the weak gravity, all the solutions belonged to the region I. 
After some critical point, the solutions for the strong gravity formed the region II. 
They had quite different shapes: the solutions of II were more compact and denser. 
We claimed that there are four regions for the solutions, but when one considers the black hole harbor, 
some of them get merged with each other. 
In fact, solutions of I and Ia merged at $r_{\rm H,crit}$ below some critical value of $b(R_{\rm in})$. 
Also, independent solutions in II merged at a $\tilde{r}_{\rm H,crit}$.  

So far, we studied the empty Minkowski-like interior, or the Schwarzschild-like black holes or the Reissner-Nordstr\"om solutions
in the normal boson shells. It became apparent that the study of the harbor solutions brings us several new insights for such gravity mediated 
solitonlike configurations. It is worth to investigate the harboring for several variants of the $Q$-balls.  
There ia a huge variety of the configurations for gauged $Q$-ball. 
It has been pointed out in \cite{Kleihaus:2010ep} that there may exist a configuration of compact boson stars inside 
the boson shells. This would lead to a funny space-time: a compact boson star surrounded by a Reissner-Nordstr\"om solution surrounded by a boson 
shell surrounded by a Reissner-Nordstr\"om solution. Though the numerical analysis will be cumbersome, certainly it should exist.  
Recently, some chain configurations of the $U(1)$ gauged $Q$-ball~\cite{Loiko:2020htk} and the boson stars~\cite{Herdeiro:2021mol} were found. 
Also, radially excited, multinode solutions of the gauged $Q$-ball model are studied in~\cite{Loginov:2020xoj, Loginov:2020wg}. 
These solutions are promising for the existence of a new harbor-type solutions. 
In our $\mathbb{C}P^N$ model, thanks to the periodicity of the compact condition, we are able to 
construct a multinode shell which also could be a harbor: the Schwarzschild or 
the Reissner-Nordstr\"om black holes are inside the shell. The study of the construction of the multinode solutions 
is almost finished and it will be reported in our next paper.

In this paper, we restrict the analysis for the solutions within the ansatz (\ref{ansatzcpn}),(\ref{ansatzgauge}), 
i.e., the solutions are composed by spherically symmetric matter profile function $f(r)$ or the gauge function $A_t(r)$ 
and the standard spherical harmonics.  Since the stress-energy tensor (\ref{stress_formal}) becomes spherically symmetric within the ansatz, 
the line element can be written as the standard Schwarzschild form (\ref{metric}).
As a result, the boson stars are spherically symmetric. 
In \cite{Nelmes:2012uf}, the authors studied the Skyrme crystal coupled with the gravitation as a model of neutron stars. 
They found that the solutions deform both isotropically and anisotropically for the strong gravity regime which allow 
the neutron star to exist with $\sim 1.90$ solar masses. 
We believe that such a solution in our model should exist, regardless of whether it is a ground state or not. 
We have to proceed our analysis in this direction. 

The stability analysis for a classical solution is important for checking validity of the symmetry imposed for finding them.  
Stability and time evolution of boson stars, the so-called study of dynamical boson stars have been done in many literatures~
\cite{Lee:1988av, Gleiser:1988rq, Seidel:1990jh, Liebling:2012fv,Alcubierre:2019qnh,Jaramillo:2020rsv}. 
Noticeable is the study for the $\ell$-boson stars, because they share some features with our model. 
For the stability of a solution, one considers small fluctuation for the field around the equilibrium. 
In \cite{Alcubierre:2019qnh,Jaramillo:2020rsv}, the authors found that for the single-field boson stars $(\ell=0)$, 
both stable and unstable branches of solutions exist for spherical perturbations. 
For multifield solutions $(\ell>0)$, the solutions remain spherically symmetric;
however, the evidence of zero modes for the nonspherical perturbation was found; 
it allows the solutions to deform with no energy loss. 
Unfortunately, the naive linear perturbation cannot apply to the compacton. 
As it was shown in \cite{Arodz:2005gz} , there is no linear regime for the compacton in terms of property of the V-shaped potential. 
That is, for the linear perturbation, one obtain a nonlinear differential equation even in the limit of 
small amplitude, which is generally not tractable.
Further, the resulting equation possesses the scaling symmetry which spoils the naive stability discussion.  
At the moment, we just conclude that the symmetry of the compactons in the $\mathbb{C}P^N$ model 
freeze-out from the phase transitions in effect of the V-shaped potential. 
It would be a big challenge to explore a new method of perturbation satisfying
the constraint of the compact support.

\begin{center}
{\bf Acknowledgments}
\end{center}
The authors would like to thank Yves Brihaye for useful advices and comments.  
We thank Universit\'e de Mons for its kind hospitality. 
We also appreciate Pawe\l~Klimas for his careful reading of this paper and many valuable advices. 
N.S. would like to acknowledge useful discussions with Yakov Shnir, Atsushi Nakamula and Kouichi Toda. 
S.Y. thanks Yukawa Institute for Theoretical Physics, Kyoto University. 
Discussions during the YITP workshop YITP-W-20-03 on ``Strings and Fields 2020'' 
were useful to complete this work. The work of N.S. was supported in part by JSPS KAKENHI Grant No.JP B20K03278(1).

	\bibliography{cpncgphase}

\end{document}